\documentstyle[pra,aps,multicol,epsfig]{revtex}
\begin{document}

\title{Delocalizing transition of multidimensional solitons in
       Bose-Einstein condensates}

\author{Bakhtiyor B. Baizakov$^\dagger$ and Mario Salerno}

\address{Dipartimento di Fisica "E.R. Caianiello"
         and Istituto Nazionale di Fisica della Materia (INFM), \\
         Universit\'a di Salerno, I-84081 Baronissi (SA),
         Italy}

\date{\today}
\maketitle

\begin{abstract}
Critical behavior of solitonic waveforms of Bose-Einstein
condensates in optical lattices (OL) has been studied in the
framework of continuous mean-field equation. In 2D and 3D OLs
bright matter-wave solitons undergo abrupt delocalization as the
strength of the OL is decreased below some critical value. Similar
delocalizing transition happens when the coefficient of
nonlinearity crosses the critical value. Contrarily, bright
solitons in 1D OLs retain their integrity over the whole range of
parameter variations. The interpretation of the phenomenon in
terms of quantum bound states in the effective potential is
proposed.
\end{abstract}


\begin{multicols}{2}

\section{Introduction}

Bose-Einstein condensates (BEC) display remarkably rich nonlinear wave
phenomena, among which solitons are particularly interesting.
To date both common types of solitons, {\it bright} and {\it dark}, are
experimentally realized in BECs, respectively, with attractive and repulsive
interatomic forces \cite{khaykovich,burger}. Theoretical studies of the
properties of matter-wave solitons has resulted in successful description of
basic features of soliton dynamics in BEC \cite{carr}.

Considerable effort is being devoted to investigation of BEC in periodic
potentials created by laser standing waves - so called optical lattices (OL).
Subjecting a BEC to a periodic potential leads to a number of interesting
phenomena, including the formation of a specific class of localized
modes called {\it gap} solitons. The definition 'gap' refers to the fact that
this type of solitons exist in the band gaps of the matter-wave spectrum.
Conceptually important point here is that, the gap solitons can be composed
of repulsive atoms, which makes them favorable against bright solitons in
continuous attractive BEC tending to collapse at high atomic densities.

After the first discussion of the possibility of bright solitons in repulsive
BEC loaded in OL \cite{potting}, substantial theoretical research was
performed aimed at understanding their development and properties \cite{OL1D}.
One of the recent advances in this direction has been the proof of the
possibility of matter-wave solitons in two- and three-dimensional (2D and 3D)
OLs. Particularly, the emergence of multi-dimensional gap solitons in
repulsive BEC arrays due to the phenomenon of modulational instability was
demonstrated in \cite{bks2002}. Different modes of stable 2D gap solitons in
OLs were shown to exist by numerical solution of the corresponding 2D
Gross-Pitaevskii equation \cite{ok2003}. Multidimensional solitons in media
governed by the self-focusing cubic nonlinear Schr\"odinger equation (NLSE)
with a periodic potential, of which an attractive BEC in OL is a particular
example, are studied in \cite{bms2003} by means of the variational
approximation and direct numerical simulations.

While have not been experimentally realized yet, bright matter-wave solitons
in OLs currently are among the intensively explored subjects in BEC. An
important aspect of OLs is that they provide a unique possibility to
investigate the behavior of localized excitations and solitons in both
continuous and discrete ends of the spectrum by adjusting the strength
of the OL. A fundamental issue that the cubic NLSE does not support stable
solitons in dimensions higher than one, while its discrete analog does,
can be addressed more effectively involving the BEC in OL.
One relevant problem - the delocalizing transition of BECs in deep 2D OLs
has recently been investigated in the framework of discrete NLSE
\cite{kalosakas}. The analysis of Ref.\cite{kalosakas} is relevant to
the tight-binding approximation, when the BECs in neighboring lattice
sites are weakly linked. This approach also relies on the possibility,
that the tunneling amplitude between OL sites can be accurately determined.
In the case of uniformly loaded OL the tunneling amplitude can be estimated
using the Wannier function basis \cite{jaksch}. However, in the case of
localized excitations involving just few lattice sites with significantly
different occupation numbers, calculation of the tunneling amplitude becomes
difficult.

The aim of this paper is twofold. From one side we present a
detailed numerical investigation of the delocalizing transition
beyond the tight-binding approximation, when the system is
described by the continuous mean-field equations. To this regard
we provide both analytical and numerical evidences for the
existence of stable multidimensional solitons in OLs, and we study
the fade and recovery of the solitonic structures as the strength
of the optical lattice or coefficient of nonlinearity
adiabatically varies in time. The major topic of interest will be
a critical behavior of solitonic waveforms in a periodic
potential, expressed as sudden disintegration of the soliton when
the system parameters attain particular values. This will be done for
BECs with both positive and negative scattering lengths. From this
study it emerges that the variational ansatz while correctly
predicting the stabilization of solitons due to the OL, it
fails to predict the existence of a delocalizing transition.

From the other side, we address to the physical mechanism
underlying the delocalizing transition. To this regard we
propose a quantum interpretation of the phenomenon, by showing
that the delocalizing transition is associated to the
disappearance of bound states from the soliton effective potential
in the underlying periodic Schr\"odinger equation. Since in one
dimension any potential well can support at least one bound state
(this is true even for infinitesimal well depths), the above
interpretation automatically implies that 1D solitons cannot
undergo delocalizing transitions. This is indeed what we find for
the 1D case. In particular we show that by decreasing the strength
of the OL or the nonlinearity in the system, the soliton becomes
very extended (in the limit of zero nonlinearity it reduces to a
Bloch state at the edge of the Brillouin zone), but always
recovers its original shape when the parameters are reversed. In
contrast, in 2D and in 3D cases, we find that there is always a
critical value of system parameters below which the recovering of
the soliton becomes impossible i.e. the soliton irreversibly
disintegrates into extended states. By approximating the soliton
effective potential with a suitable solvable potential and by
adopting a variational ansatz description for the soliton, we show
that the above {\it bound state interpretation} leads to an
analytical expression for the occurrence of the delocalizing
transition which is in good agreement with direct numerical
simulations of the Gross-Pitaevskii equation, this providing
support for our approach.

We remark that the proposed mechanism for the delocalizing
transition is general and  expected to be valid also for
multidimensional intrinsic localized modes or discrete breathes of
nonlinear lattices, as well as for multidimensional soliton
solutions of the NLSE with nonlocal interactions.

Finally, we have explored the role of dimensionality by comparison of results
for 2D and 3D OLs. The advantage of the present approach is that, all the
system variables are readily connected to actual physical parameters
(e.g. the OL strength to the laser intensity and the coefficient of
nonlinearity to the atomic scattering length, etc.), which makes the
experimental verification more feasible.

The paper is organized as follows. In Sec. II we introduce the model and
derive variational equations for soliton parameters. Then we check the
validity of VA equations by comparing with direct numerical solution
of the Gross-Pitaevskii equation. In Sec. III we propose a physical
mechanism associating the existence of solitons and their delocalizing
transition with quantum bound states in the effective potential. The
existence region for 2D solitons is presented. In Sec. IV we discuss
the possibility of experimental observation of the delocalizing transition
of solitons. Finally, in Sec. V we conclude the results of this study.

\section{The existence of solitons in periodic potentials}

The existence, stability and some other properties of solitons in
1D periodic potentials are explored in the context of repulsive
BEC in OLs \cite{potting,OL1D}. Solitons in self-focusing NLSE
with a 1D periodic potential, with relevance to particular
problems of nonlinear optics is considered in \cite{malomed99}.
While the subject is well understood in 1D case, multidimensional
solitons in periodic potentials are in early stage of research
\cite{bks2002,ok2003,bms2003}. Below we analyze the existence of
multidimensional solitons in periodic potentials in the framework
of the variational approximation (VA) for both attractive and
repulsive interactions. The results will then be  compared with
numerical simulations of the full problem. We remark that the VA
analysis based on a Gaussian waveform was recently considered in
Ref. \cite{bms2003} for the case of 2D BEC with attractive
interactions. In this case it was shown that bright solitons are
well approximated by Gaussian profiles, this being particularly
true when all the matter is confined into a single cell of the OL
(single cell solitons). In the case of BEC with repulsive
interactions, however, satellite peaks appear in the soliton
waveform and the VA based on a Gaussian ansatz leads to wrong
results. In this case we will show that a Fraunhofer diffraction
pattern is a better  ansatz for the soliton shape, leading to VA
equations which are in good agreement with numerical simulations.

\subsection{Basic equation}

The analysis will be based on the mean-field Gross-Pitaevskii equation
(GPE) for the macroscopic wave function of the condensate $\psi(\bar r, t)$,
which describes the properties of a dilute gas near-zero temperature
BEC \cite{dalfovo}
\begin{equation} \label{gpe}
  i\hbar \frac{\partial \psi}{\partial t} = \bigl[-\frac{\hbar^2}{2 m}
  \nabla^2 + V_{ext}(\bar r,t) +g |\psi|^2 \bigr] \psi,
\end{equation}
where $m$ is the atomic mass, $g$ is the nonlinear coupling parameter related
to the $s$-wave scattering length $a_s$ through $g=4 \pi \hbar^2 a_s/m$,
$V_{ext}(\bar r,t)$ is the external potential, generally consisting of
harmonic confinement plus periodic potential of the OL
\begin{equation} \label{Vext}
  V_{ext}(\bar r,t) = \frac{m}{2}\omega_i^2 r_i^2 +
  \varepsilon \cos(2 k_L r_i),
\end{equation}
where $r_i=x,y,z$ and sum over $i$ is assumed. We shall be interested in
the behavior of localized excitations, occupying few central unit cells of
the OL, when the effect of confining parabolic potential is not essential.
Therefore, in what follows we assume the presence of the OL only.
For convenience we rescale the Eq.(\ref{gpe}) by introducing the
dimensionless variables $r_i \rightarrow k_L r_i$,
$t \rightarrow (\hbar k_L^2)/(2 m) \cdot t$, $\varepsilon \rightarrow
\varepsilon/E_r$, $\psi \rightarrow \psi/\sqrt{n_0}$. Where
$k_L=2 \pi/\lambda$, $E_r = (\hbar^2 k_L^2)/(2 m)$,
$\lambda, k_L$ being the laser wavelength and wave vector correspondingly,
$n_0$ is the density of BEC. We also designate the coefficient of
nonlinearity as $\chi = 8 \pi n_0 a_s/k_L^2$. Then we have Eq.(\ref{gpe})
in the form
\begin{equation} \label{gpe1}
  i \frac{\partial \psi}{\partial t} = - \nabla^2 \psi + V_{ol}(r,t) \psi +
  \chi |\psi|^2 \psi,
\end{equation}
where $V_{ol}(r,t)=\varepsilon [\cos(2 x)+\cos(2 y)+\cos(2 z)]$ is the OL
potential.

\subsection{Variational approach}

To deal with compact expressions, we explicitly consider the 2D
case, since the extension to 3D is straightforward.
We look for the stationary solution of Eq.(\ref{gpe1}) in the form
\begin{equation}
  \psi(x,y,t) = U(x,y) \exp(-i \mu t),
\end{equation}
where $\mu$ is the chemical potential. Then the following equation
determines the time-independent solution $U(x,y)$
\begin{equation} \label{stat}
  \nabla^2 U + \mu U + V_{ol}(x,y) U + \chi U^3 = 0.
\end{equation}
Following the standard procedure of the VA \cite{anderson1983,malomed2002},
we consider the effective Lagrangian
\begin{equation} \label{lagrangian}
  L = \frac{1}{2}\int_{\infty}^{\infty} \bigl[(\nabla U)^2 -
  \mu U^2 - V_{ol}(x,y) U^2 - \frac{1}{2} \chi U^4 \bigr] dx dy
\end{equation}
which generates the Eq.(\ref{stat}) by minimization $\delta L = 0$.
The next step of the variational approach involves the choice
of a suitable trial function - {\it ansatz} for the solution. As we
shall see later, solitonic waveforms in attractive and repulsive
BECs in the periodic potential have different spatial features.
This implies the choice of different trial functions, which we
shall consider separately.

\subsubsection{Attractive BEC}

\begin{figure}[htb]
\centerline{\includegraphics[width=8.0cm,height=6.0cm,clip]{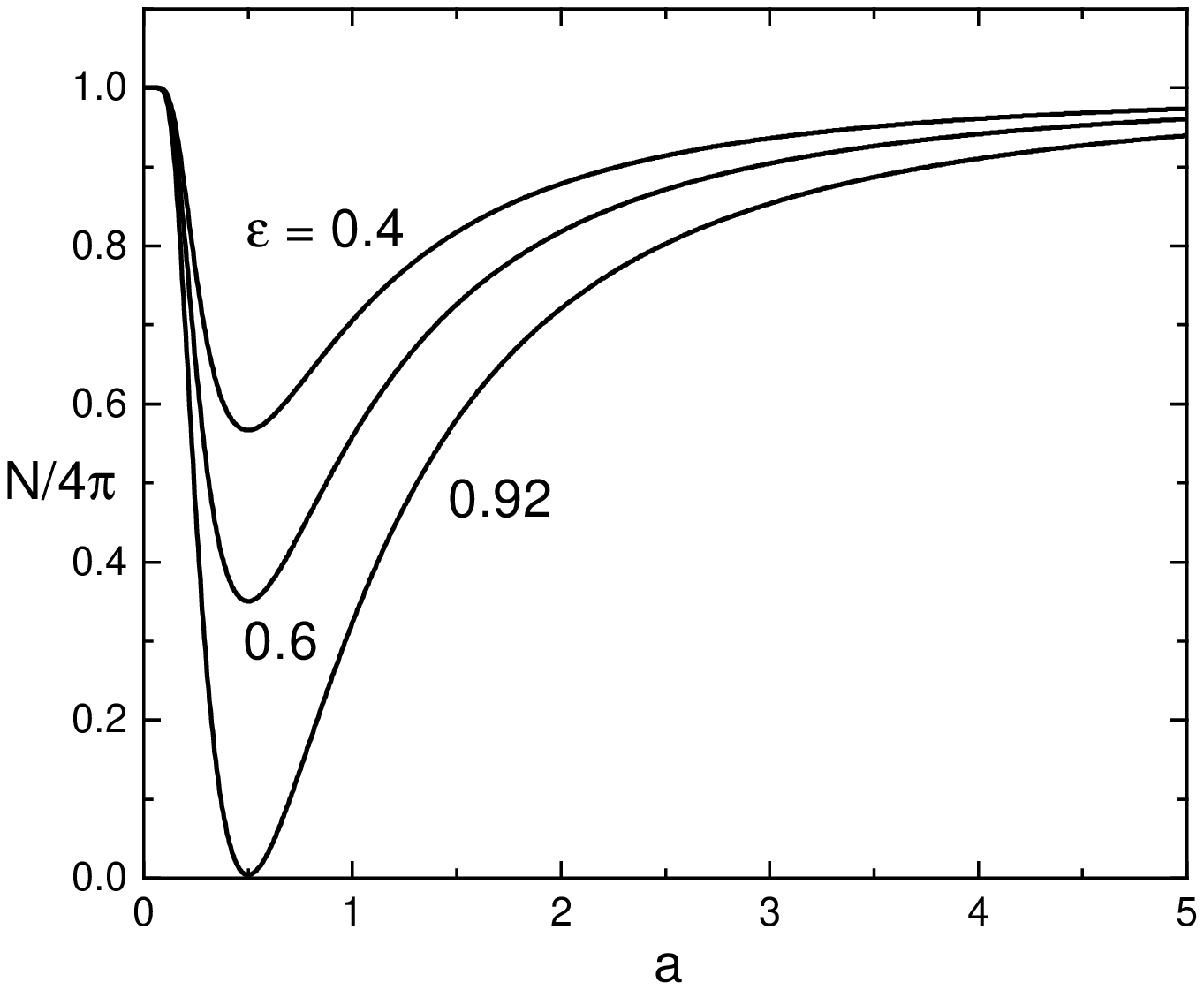}}
\caption{The Eqs.(\ref{VA1}) predict two localized solutions for a given
norm $N$, one of which ($a>1/2$) is stable and the other ($a<1/2$)
is unstable.}
\label{f1}
\end{figure}

The nature of interatomic forces in BEC can be attractive or repulsive,
which leads to very different properties of corresponding BECs \cite{dalfovo}.
The attractive case corresponds to the negative sign of the s-wave scattering
length $a_s$, and the underlying GPE has a self-focusing nonlinearity
($\chi < 0$). The collapse of BEC at some critical number of
atoms $N \sim 1400$ is the main consequence of the self-focusing
nonlinearity \cite{dalfovo}. The presence of a periodic potential, however,
significantly changes the situation and leads to stable localized states.
In moderately strong and weak periodic potentials (which is the
case at delocalizing transition point considered below) multidimensional
solitons have a single cell structure and are well described by the VA with
Gaussian ansatz \cite{bms2003}
\begin{equation} \label{ansatz1}
  U(x,y) = A \, \exp[-\frac{a}{2}(x^2+y^2)],
\end{equation}
where the amplitude $A$ and width parameter $a$ are fixed by the
total number of atoms in the condensate
$N = \int_{-\infty}^{\infty} |U(x,y)|^2 dx dy = \pi A^2/a$. Variational
equations are derived by substituting the ansatz (\ref{ansatz1})
into the effective Lagrangian (\ref{lagrangian}),
performing the spatial integration and subsequent minimization with respect
to variables $A$ and $a$.
The equations associate the parameters of the soliton with the total
number of atoms, strength of the OL, and chemical potential of the BEC
\begin{equation} \label{VA1}
  N = \frac{4\pi}{\chi} \Bigl(1-\frac{2\varepsilon}{a^2} e^{-1/a}\Bigr), \quad
  \mu = \frac{2\varepsilon}{a} (2-a) e^{-1/a} -a.
\end{equation}
A noteworthy property of Eq.(\ref{VA1}) is that, for a given
strength of the OL $\varepsilon$ there exists a minimal norm
$N_{min} = \frac{4\pi}{\chi}(1-8\varepsilon e^{-2})$ attained at
$a=1/2$, below which the localized solutions do not exist, as
illustrated in Fig.\ref{f1}. The threshold vanishes for stronger
OLs, $\varepsilon > \varepsilon_{cr} = e^2/8 = 0.92$. The
existence of solitons in 2D attractive BEC from above is limited
by the onset of collapse at the critical norm $N_{col} = 4\pi$
predicted by VA \cite{abd2003} (the exact value being $N_{col} =
11.6993$ \cite{berge}).
The VA predicts two localized states with different widths for a given norm
$N_{min} < N < N_{col}$. The stability of these two solutions can be inferred
from the Vakhitov-Kolokolov criterion \cite{VK}. According to this criterion,
the branch of solutions with $a>1/2$ has the negative slope $dN/d\mu < 0$,
and therefore is stable. The other branch ($a<1/2$) appears to be unstable.

\subsubsection{Repulsive BEC}

The main feature of a soliton of repulsive BEC ($\chi>0$) in a periodic
potential is that, it is composed of a central peak and symmetrically
spaced satellites residing in neighboring cells of the OL. The issue of
vital importance for VA is the adoption of a suitable ansatz. A Gaussian
waveform obviously is not satisfactory in this case since it doesn't
take into account the composite shape of solitons. We introduce the following
ansatz
\begin{equation} \label{ansatz2}
  U(x,y) = A \cdot \frac{\sin(ax)}{ax} \cdot \frac{\sin(ay)}{ay},
\end{equation}
which accounts for the satellites, and has the advantage of
yielding simple VA equations. Analytical transformations similar
to the previous case leads to following set of equations
\begin{equation} \label{VA2}
  N = \frac{3\pi^2}{\chi} \Bigl( \frac{3\varepsilon}{2a^3} -1 \Bigr), \quad
  \mu=\frac{2\varepsilon}{a}(2-a) - \frac{2}{3} a^2.
\end{equation}
The relation connecting the norm and soliton parameters in this case
is $N=\pi^2 A^2/a^2$, which yields in combination with Eqs.(\ref{VA2})
\begin{equation} \label{aeps}
  A = \Bigl[ \frac{9 N^{3/2} \varepsilon}{2\pi(N \chi+3\pi^2)} \Bigr]^{1/3}.
\end{equation}
Unlike the attractive case, the ansatz Eq.(\ref{ansatz2}) yields a single
set of soliton parameters $a, A$ for a given norm $N$ and strength of
the OL $\varepsilon$. In the next subsection we compare the prediction of
the VA with direct numerical solution of the GPE (\ref{gpe1}).

\subsection{Comparison of VA and PDE simulations}

Validity of the VA is usually verified by comparison with the
results of direct numerical solution of the underlying PDE. Below
we solve the Eq.(\ref{gpe1}) by split-step procedure using the
multi-dimensional fast Fourier transform \cite{numrecipes} on the
grids of 512, 256x256 and 128x128x128, respectively in 1D, 2D and
3D cases. The domain size and time step are $x,y,z \in [-4\pi,
4\pi]$, $\delta t = 0.001$. The essential point is the use of
absorption on the domain boundaries to prevent re-entering of
linear waves emitted by the soliton during its formation (or under
perturbations) into the integration area.

The comparison between VA and PDE simulations can be done in
different ways. We consider the situations more relevant to
experiments, namely, we change the strength of the OL
$\varepsilon$ or the coefficient of nonlinearity $\chi$ (which is
equivalent to changing the norm $N$ at constant $\chi$), and
follow the evolution of the soliton's amplitude according to PDE
(\ref{gpe1}). Then we correlate the numerically obtained
dependence $A(\varepsilon)$ or $A(N)$ with analytical prediction
of VA Eqs.(\ref{VA1},\ref{VA2}). Below we perform the comparisons
for 2D case.

\subsubsection{Attractive case}

\begin{figure}[htb]
\centerline{\includegraphics[width=8.0cm,height=2.0cm,clip]{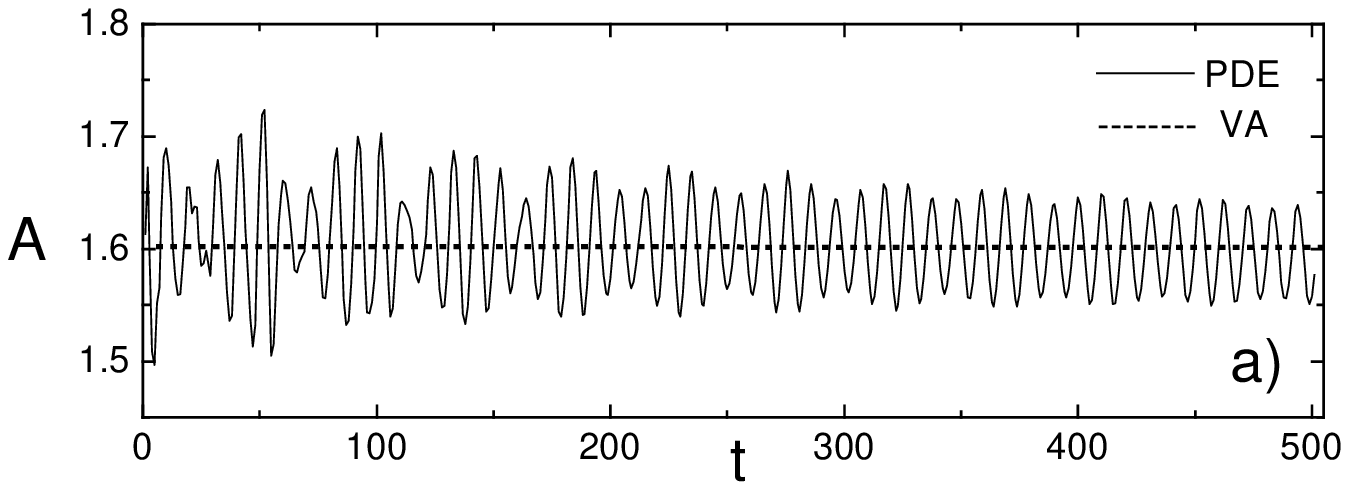}}
\centerline{\includegraphics[width=8.0cm,height=5.0cm,clip]{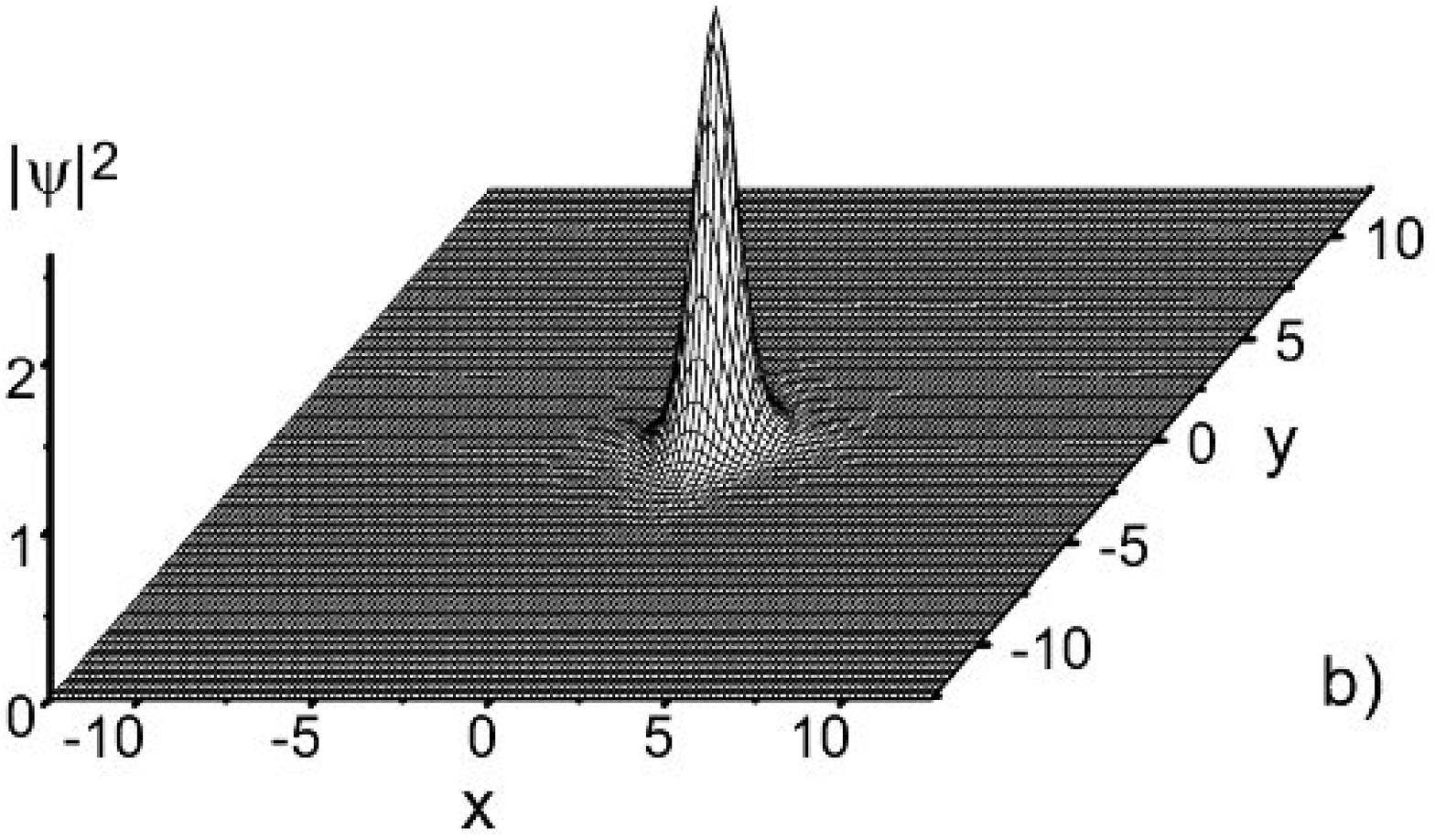}}
\caption{The Gaussian waveform Eq.(\ref{ansatz1}) with parameters
prescribed by VA $A=1.6, \ a=1.3, \ N=2\pi$,
inserted as initial condition to Eq.(\ref{gpe1}) evolves into a stable 2D
soliton, performing transient oscillations. The strength of OL is
$\varepsilon = 0.92$. a) Comparison VA vs. PDE, b) stable 2D soliton of
attractive BEC.}
\label{f2}
\end{figure}

At first we need to generate a stable 2D soliton of Eq.(\ref{gpe1}). For this
we insert the Gaussian waveform (\ref{ansatz1}) with parameters specified
by VA (see Fig.\ref{f1}) into the Eq.(\ref{gpe1}) as the initial
condition. The waveform undergoes some evolution and attains the stable
state as shown in Fig.\ref{f2}.
The stability of multidimensional solitons, being a very important issue,
is worth of special investigation. Here we refer to our numerical tests,
which show that a small deformation of the soliton
causes weak oscillations around the stable state. After sufficiently
long time the waveform relaxes back to its original shape proving that
the stable state is a fixed point, i.e. the soliton is linearly stable.
The tests also involve the long-time propagation of a fundamental soliton
with PDE for a period much exceeding the characteristic time of the problem.

\begin{figure}[htb]
\centerline{\includegraphics[width=8.0cm,height=6.0cm,clip]{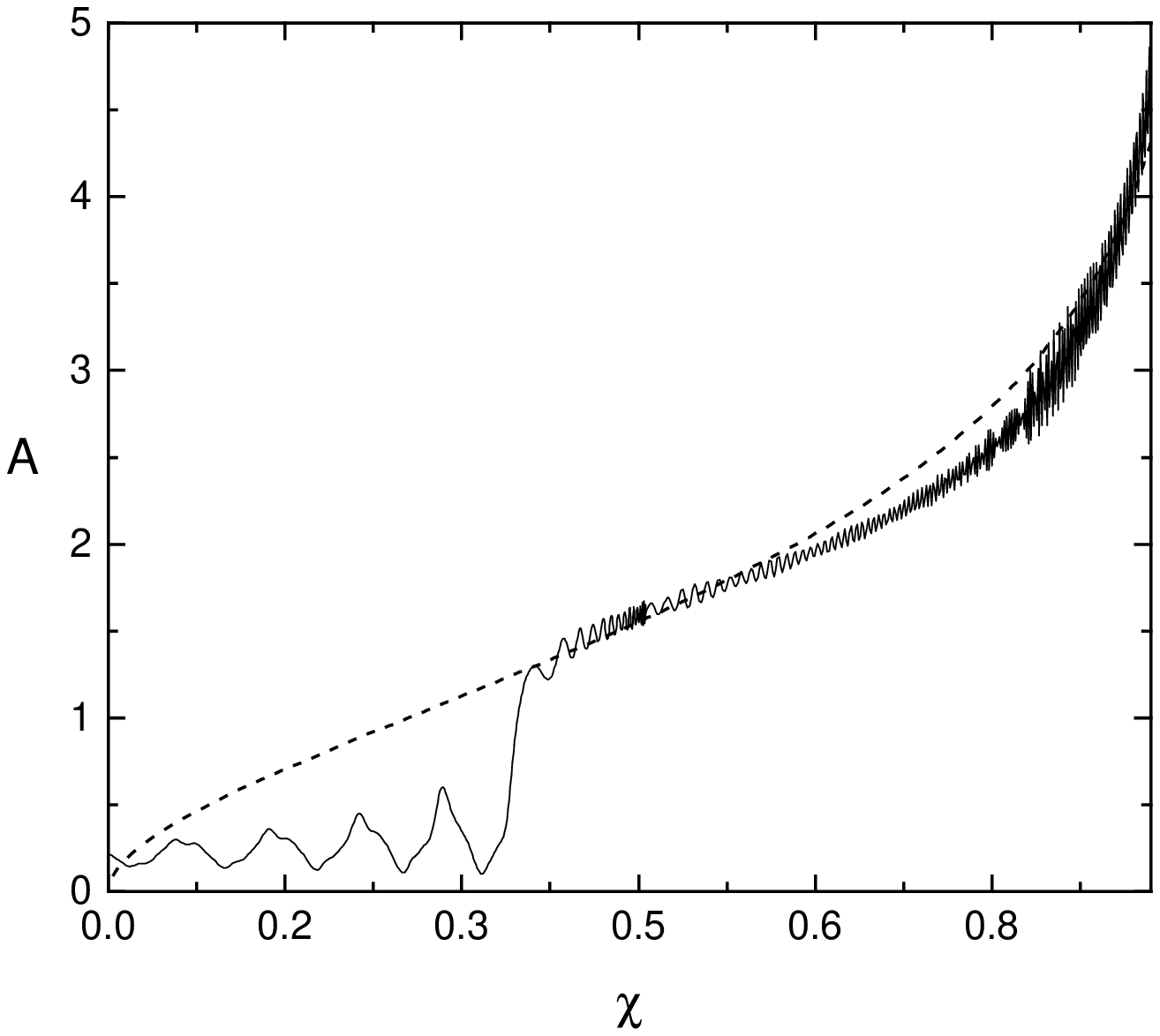}}
\caption{The amplitude of a 2D soliton of attractive BEC as a function
of the coefficient of nonlinearity, obtained by numerical solution of
the GPE (\ref{gpe}) (solid line), and as predicted by VA Eq.(\ref{VA1})
(dashed line). VA adequately describes the dependence $A(\chi)$ above the
delocalizing transition point at $\chi = 0.38$. }
\label{f3}
\end{figure}

After the stable 2D soliton is created, we use it as initial condition for
Eq.(\ref{gpe1}) and follow the further evolution under time-dependent
parameters $\varepsilon(t)$ or $\chi(t)$.
An important point to be stressed here is the condition of
adiabaticity in variation of parameters. In the present context, the
adiabaticity refers to the situation, when the variation of parameters
does not excite collective modes of the condensate, which means that
the ramp times should satisfy the condition $t_{ramp} > \hbar/\mu$.
In our dimensionless units the above condition implies $t_{ramp} > 50$
for a typical condensate with chemical potential of $\mu = 200$ Hz. The
soliton's amplitude as a function of the coefficient of nonlinearity,
obtained from PDE simulations and compared with the prediction of VA is
presented in Fig.\ref{f3}.
An important remark concerning the Fig. {\ref{f3}} is that the VA
fails to account for the delocalization of a soliton occurring
when the norm drops below some critical value ($\chi = 0.38$, for
$\varepsilon = 0.92$), which is manifested as a rapid spreading of
the waveform.

\subsubsection{Repulsive case}

\begin{figure}[htb]
\centerline{\includegraphics[width=8.0cm,height=2.0cm,clip]{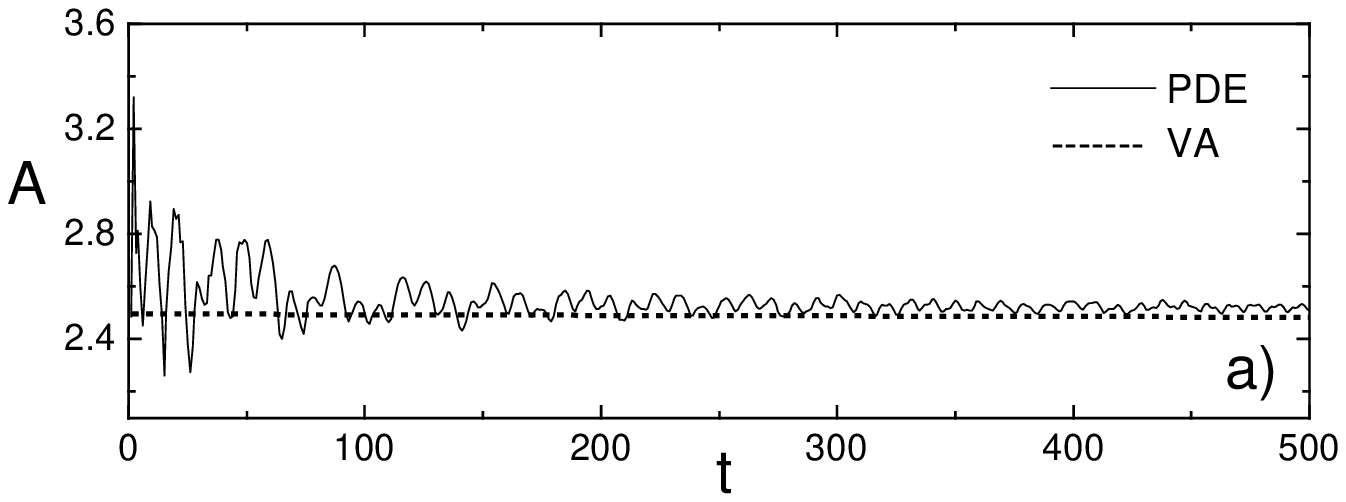}}
\centerline{\includegraphics[width=8.0cm,height=5.0cm,clip]{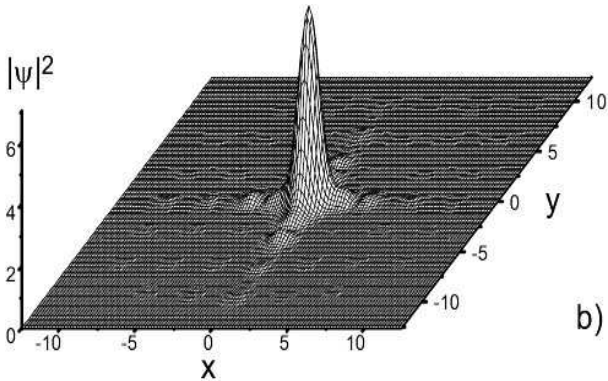}}
\caption{Formation of a 2D soliton from the initial state given by
Eq.(\ref{ansatz2}) with $A=2.5, \ a=1.1$ in the OL of strength
$\varepsilon = 4.0$. a) Transient oscillations of the amplitude,
b) stable 2D soliton of repulsive BEC.  The norm has decreased by
50 \% during the formation of a stable soliton. }
\label{f4}
\end{figure}

\begin{figure}[htb]
\centerline{\includegraphics[width=8.0cm,height=4.0cm,clip]{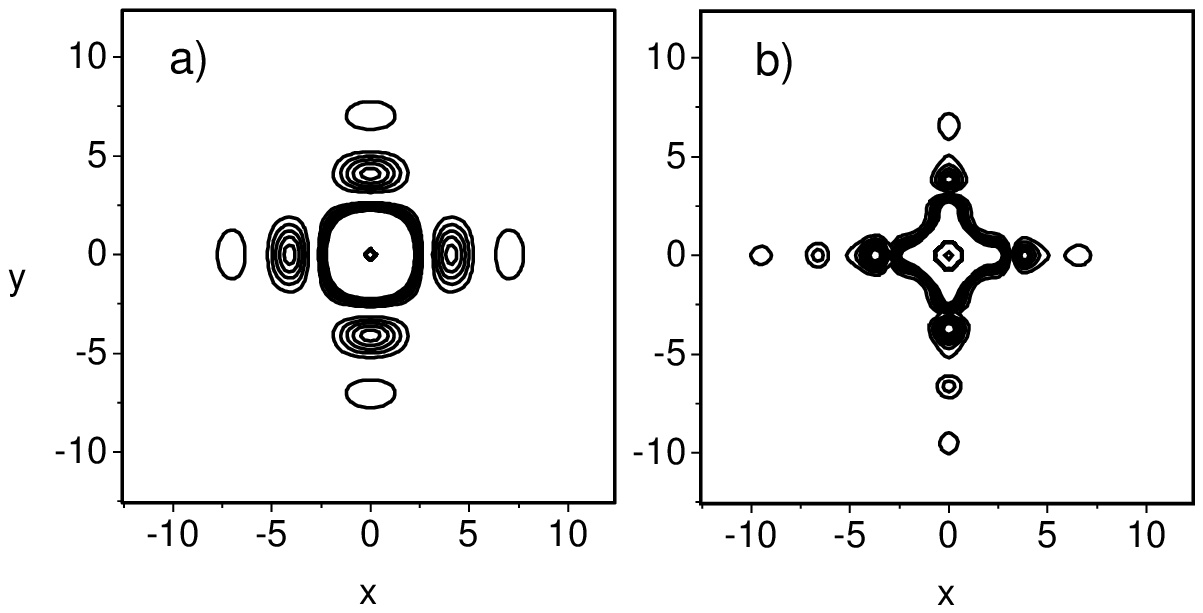}}
\caption{Contour plots of the ansatz Eq.(\ref{ansatz2}) (a),
and the stable 2D soliton (b) of repulsive BEC. }
\label{f5}
\end{figure}

\begin{figure}[htb]
\centerline{\includegraphics[width=8.0cm,height=6.0cm,clip]{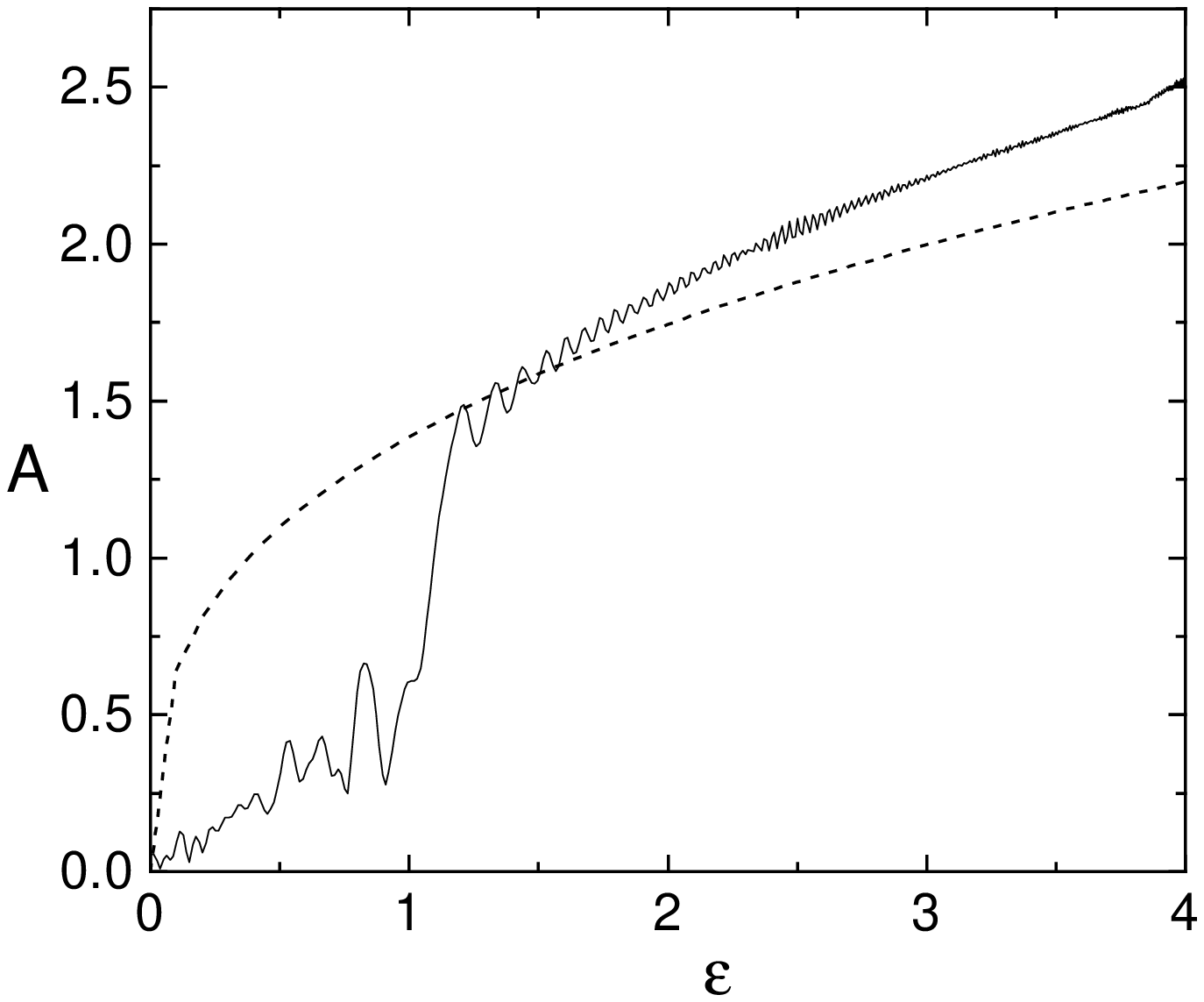}}
\caption{The amplitude of a 2D soliton of repulsive BEC as a function
of the strength of the periodic potential. Solid line - numerical solution
of Eq.(\ref{gpe1}), dashed line - prediction of VA Eq.(\ref{aeps}).}
\label{f6}
\end{figure}

Similarly to the previous case, we generate a stable 2D soliton of repulsive
BEC inserting the waveform Eq.(\ref{ansatz2}) with $A=2.5, \ a=1.1$ as
initial condition for the Eq.(\ref{gpe1}) with $\varepsilon = 4.0$ and
$\chi = 1$. After some transition period a stable soliton is formed as
shown in Fig.\ref{f4}. In this case the oscillations of the amplitude
around the stable state is less pronounced due to repulsive nature of
the condensate.
The fact that the ansatz (\ref{ansatz2}) properly accounts for the satellites
can be seen from the comparison of contour plots in Fig.\ref{f5}.
Fig.\ref{f6} illustrates the soliton's amplitude as a function of the
strength of the OL given by VA Eq.(\ref{aeps}), and obtained by direct
numerical solution of Eq.(\ref{gpe1}).
Here again, as in the attractive case, we see that the VA fails to predict
the delocalizing transition (at $\varepsilon \simeq 1.2$). Below we
consider in detail the delocalizing transition of multidimensional solitons
in periodic potentials.

\section{Delocalizing transition}

\begin{figure}[htb]
\centerline{\includegraphics[width=8cm,height=4cm,clip]{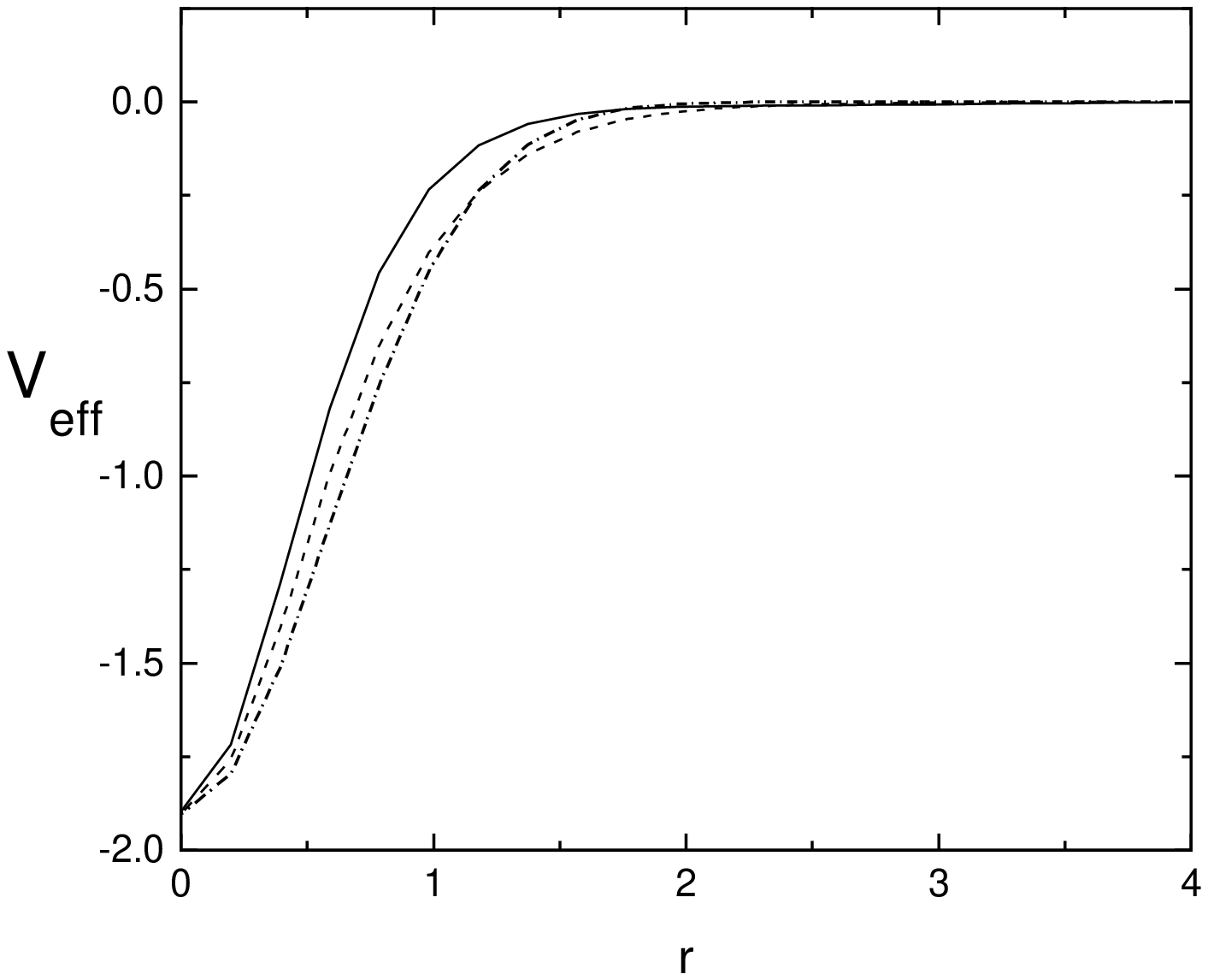}}
\vspace*{0.5cm}
\caption{The effective potential of a 2D soliton of attractive BEC
at the delocalizing transition point, obtained from numerical solution
of the GPE (\ref{gpe1}). The dashed line refers to the P\"oschl-Teller
potential in Eq. (\ref{pt}), while the dash-dotted line is the Gaussian
approximation Eq.(\ref{ansatz1}) with parameters $A_d=1.38$, $a_d=1.25$,
corresponding to the delocalizing transition point of Fig.\ref{f3}
at $\chi = 0.38$ for $\varepsilon=0.92$.}
\label{f7}
\end{figure}

The delocalizing transition of solitons is manifested as irreversible
transformation of the localized waveform to the extended state.
The transition can be induced by decreasing of the strength of the periodic
potential, or reducing the coefficient of nonlinearity in Eq.(\ref{gpe1}),
below some critical value. As pointed out in the previous section, the
delocalizing transition of solitons is missed in VA description. In this
section we present an interpretation of the phenomenon, involving the
quantum bound states in the effective potential created by the soliton.
For this we assume the condensate to have the form of a
single cell soliton, and consider the VA equations valid until the
delocalizing transition occurs. An example of this is shown in
Fig. \ref{f3} for the case of attractive interactions.
The underlying idea is to consider the GPE (\ref{gpe1}) as a linear
Schr\"odinger equation with the effective potential
\begin{equation} \label{effpot}
  V_{eff} = V_{ol} + \chi |\psi|^2,
\end{equation}
where $V_{ol}$ is the periodic potential of the OL,
and the second term represents the contribution of the soliton.
Finding the stationary solution of the Schr\"odinger equation with
the potential (\ref{effpot}) is similar to the self-consistent
Hartree-Fock approximation of quantum mechanics. From this point
of view it is clear that the existence of matter-wave solitons is
linked to the existence of quantum bound states in the effective
potential (\ref{effpot}). We remark that the periodic part
$V_{ol}$ can be eliminated from the effective potential. This is
particularly connected to the possibility, that the GPE
(\ref{gpe}) with a periodic potential can be reduced to the
standard NLSE in the effective mass formalism \cite{steel}. In
other words, the contribution of $V_{ol}$ in the linear
Schr\"odinger problem can be accounted in terms of an effective
mass description for the extended states of the linear problem.
These states have positive or negative effective masses depending
on their interaction being, respectively, repulsive or attractive
(they resemble electrons or holes of usual solids). In both cases
we reveal that the effective potential reduces only to the soliton
part with renormalized coefficient of nonlinearity $\chi$
\cite{steel}
\begin{equation} \label{Veff}
  V_{eff}=-|\chi| |\psi (x,y)|^2,
\end{equation}
where $\psi(x,y)$ denotes the solution of the nonlinear problem in
the presence of OL, evaluated at the delocalizing
transition point. Notice that the effective potential is always
negative, i.e. the soliton acts always as a potential well in the
corresponding Schr\"odinger problem (in the repulsive case the
sign of the potential is reversed by the negative effective mass).
The problem then is to investigate the existence of states with
negative energy of the following Schr\"odinger equation
\begin{equation} \label{schro}
  (-\nabla ^2 + V_{eff}-E) \phi(x,y)=0.
\end{equation}
For this one can resort to numerical schemes for Eq. (\ref{schro})
to find that at the delocalization point the effective potential
has the critical strength to support just a single bound state.
In the following, however, we shall provide an
analytical estimation of the number of bound states employing the
VA solutions discussed in the previous sections
and approximating the effective potential with a solvable
potential for which this number is exactly known. To this end we
use the following P\"oschl-Teller potential \cite{poschl}
\begin{equation} \label{pt}
  V_{PT}(r)=- \frac{\chi A^2}{\cosh(r {\sqrt{ 2 a \ln 2}})^2},
\end{equation}
to approximate $V_{eff}$ at the delocalization transition. The
parametric form of  $V_{PT}$ is fixed so that $V_{PT}$ has the
same amplitude and the same integral value:
$2 \pi \int_0^\infty r V_{eff}(r) dr$,  as for $V_{eff}$.
In Fig.\ref{f7} $V_{eff}$ and its P\"oschl-Teller approximation are depicted
for the parameter values corresponding to the delocalization transition of
Fig.\ref{f3}.
The number of bound states of the P\"oschl-Teller potential is
given by the integer part of $n$ \cite{flugge}
\begin{equation}\label{n}
  n=\frac{1}4 (\sqrt{1+\frac{2 \chi A^2}{a \ln 2}}+1).
\end{equation}
This equation can be reduced to a more convenient form by using the
expressions for the amplitude $A$ and the width $a$ of the soliton
as obtained from VA. At the delocalizing transition $n = 1$
so that, after substitution $A^2/a = N/\pi$, the Eq.(\ref{n}) reduces to
$\frac{N \chi}{4 \pi}=\ln 2$. This means that there exists a critical value
$N_{cr} = 4 \pi \ln 2/\chi$ for the number of atoms, below which the soliton
disintegrates (notice that this delocalization threshold is a factor
of $\ln 2$ smaller than that of unstable 2D Townes soliton \cite{berge}).
Since $N_{cr}$ is related to the amplitude and width $A_d, a_d$,
of the soliton at the delocalization point by the variational
equations, we can express the delocalization condition in terms of
$\varepsilon, a_d, A_d,$ as
\begin{equation}\label{nva}
  \varepsilon=\frac{1}{2} a_d^2 (1-\ln{2}) e^{1/a_d}, \quad
  \frac{A_d^2 \chi}{4 a_d}=\ln 2.
\end{equation}
In the following we shall compare these expressions for the
delocalization transition with the results of direct PDE simulations.

\subsection{1D optical lattice}

Since the 1D Schr\"odinger equation has bound state solution in
any confining potential (even of infinitesimal depth), this
implies that the delocalizing transition of bright solitons cannot
occur in 1D case. It is interesting to start with consideration of
this situation, since it provides more insight into the general
problem.

\begin{figure}[htb]
\centerline{\includegraphics[width=8.0cm,height=5.0cm,clip]{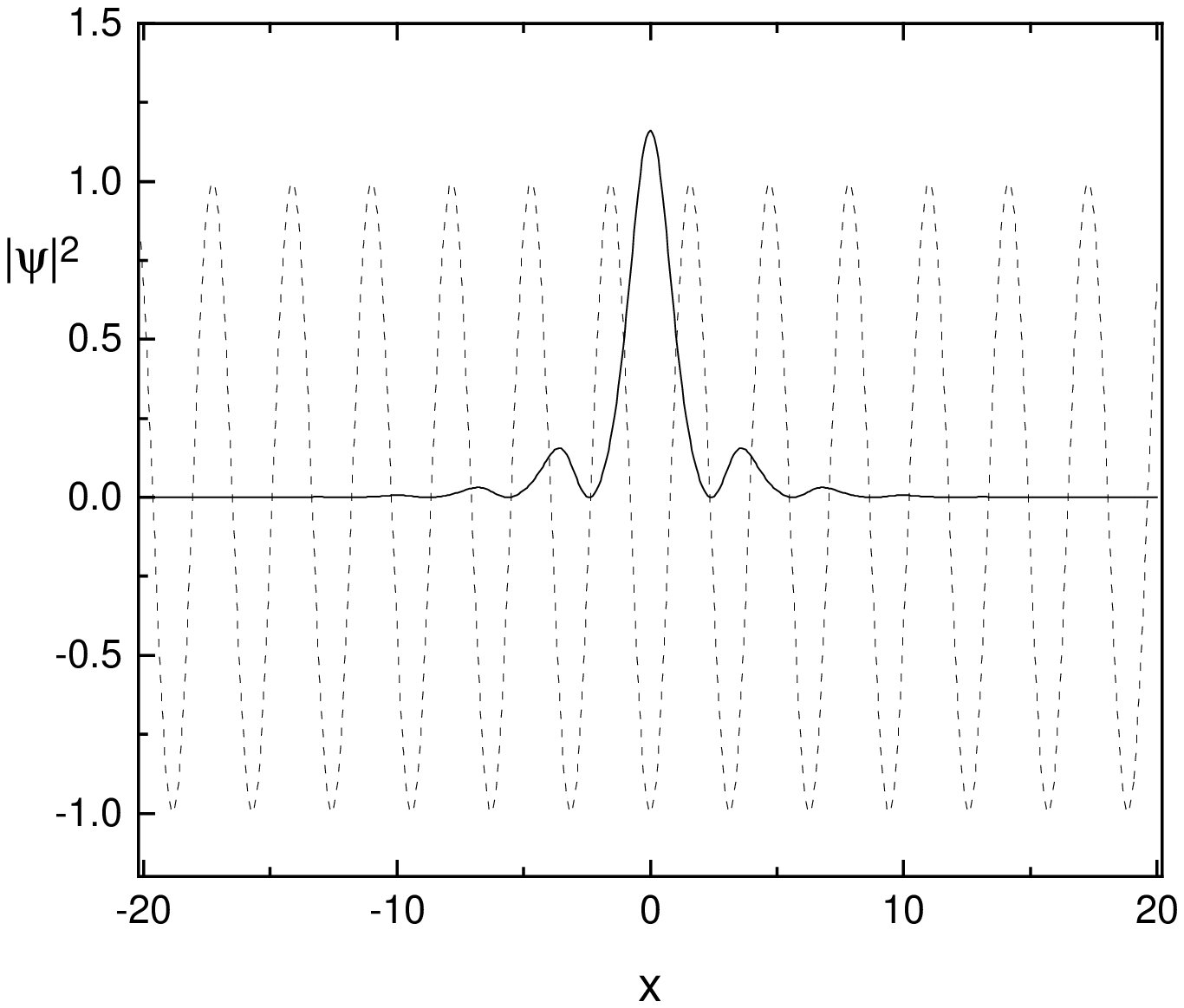}}
\caption{The initial waveform corresponding to stationary
solution of Eq.(\ref{gpe1}) with parameters $\varepsilon = 1, \ \chi=1.$
Dashed line represents 1D OL potential $V(x)=\varepsilon \cos(2 x).$}
\label{f8}
\end{figure}

\begin{figure}[htb]
\centerline{\includegraphics[width=8.0cm,height=6.0cm,clip]{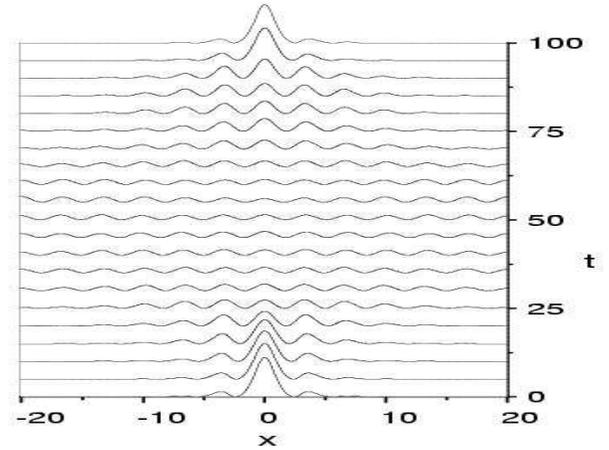}}
\caption{When the strength of the OL $\varepsilon_0=1$ is
decreased to zero at $t=50$ and then increased back to its initial value,
the 1D gap soliton of repulsive BEC fully restores its original form.}
\label{f9}
\end{figure}

The dynamics of the condensate is governed by 1D version of the
GPE (\ref{gpe1}). The stationary localized solution to this equation can
be found by standard numerical relaxation procedure.
One of such gap soliton solutions for repulsive case ($\chi>0$) is
presented in Fig.\ref{f8}
Now we explore the behavior of 1D soliton under adiabatic variation of
the strength of OL $\varepsilon(t)$ or coefficient of nonlinearity
$\chi(t)$, using the above solution (Fig.\ref{f8}) as initial condition
for time-dependent GPE (\ref{gpe1}). We assume the simplest linear function
for variation of the amplitude $\varepsilon(t) = \varepsilon_0 \cdot f(t)$
and coefficient of nonlinearity $\chi(t) =\chi_0 \cdot f(t) $, with
\begin{equation} \label{ft}
f(t) = \left \{
\begin{array}{lcl}
1-2 \alpha (t/t_{end}), & \mbox{if} & 0 \leq t \leq 0.5 \cdot t_{end} \\
1-2 \alpha (t_{end}-t)/t_{end}, & \mbox{if} & 0.5 \cdot t_{end} < t \leq t_{end} \\
\end{array}
\right.
\end{equation}
In Fig.\ref{f9} we report the transformation of the solitonic state into
the extended one, and restoring its original shape as the strength of the
OL is adiabatically decreased to zero at $t=50$, and then increased back to
initial value at $t=100$.

\begin{figure}[htb]
\centerline{\includegraphics[width=8.0cm,height=4.0cm,clip]{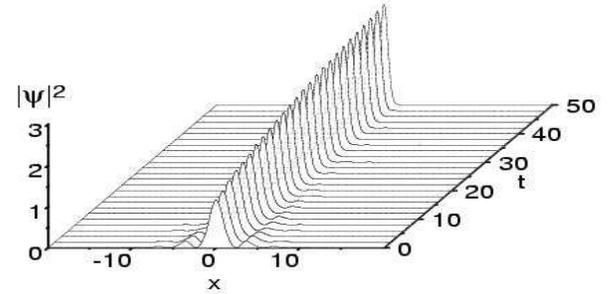}}
\caption{All the BEC matter is pulled into the central unit cell
as the strength of the OL is linearly increased from
$\varepsilon=1$ to $\varepsilon=5$ during $t  = 50 $. }
\label{f10}
\end{figure}

\begin{figure}[htb]
\centerline{\includegraphics[width=8.0cm,height=4.0cm,clip]{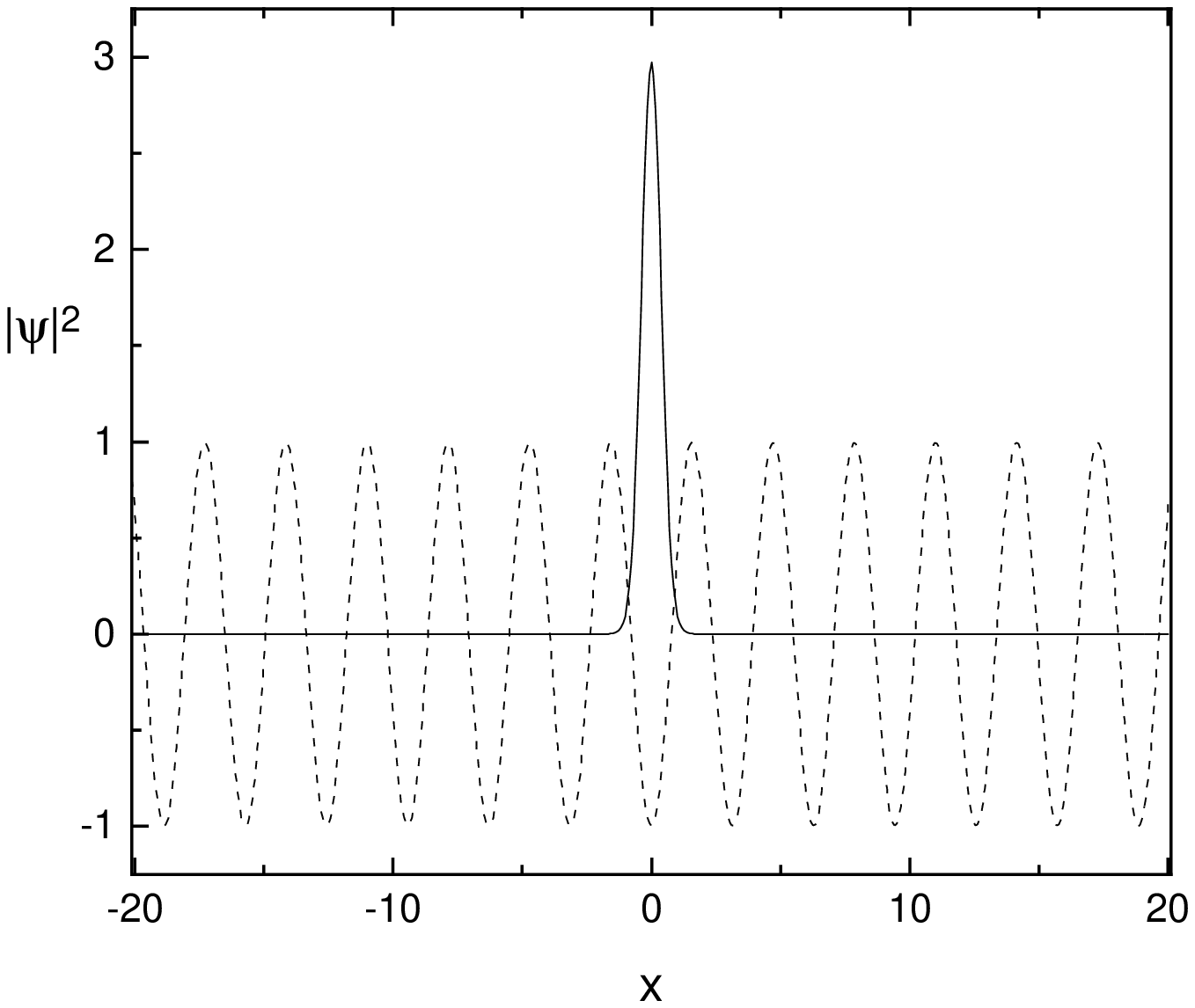}}
\caption{A 1D gap soliton of Fig.\ref{f8} contracts into the central
unit cell when the strength of the OL is increased from
$\varepsilon =1$ to $\varepsilon = 5$. Dashed line represents
$\cos(2 x)$}
\label{f11}
\end{figure}

The opposite situation also reveals the solitarity of the waveform
(Fig.\ref{f8}). Specifically, when the strength of the OL
is increased enough the BEC matter is entirely pulled into the
central unit cell (Figs. \ref{f10}, \ref{f11}).

The following Fig.\ref{f11} represents the waveform in OL potential
at $t=50$.

\begin{figure}[htb]
\centerline{\includegraphics[width=8.0cm,height=6cm,clip]{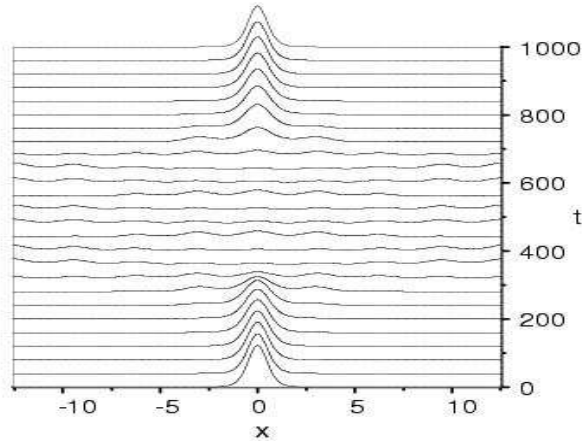}}
\caption{Soliton of attractive BEC in 1D OL of strength
$\varepsilon = 1$ retains its integrity while the coefficient of
nonlinearity is decreased to zero at $t=500$, and then increased
back to the original value $\chi=-1$ at $t=1000$.} \label{f12}
\end{figure}

In the case of attractive BEC ($\chi < 0$), the stationary solution of the
GPE (\ref{gpe1}) can be found similarly to above, by relaxation method.
Below we consider the solution corresponding to the strength of the OL
$\varepsilon = 1.0$, and the soliton initially fits a single cell of
the periodic potential. Then we use this solution as initial condition for
GPE (\ref{gpe1}) and observe the spreading and contracting of the waveform as
the coefficient of nonlinearity is decreased to zero and increased back to
its original value, as displayed in Fig.\ref{f12}

To conclude this subsection we note, that the bright solitons in
1D periodic potentials do not exhibit the delocalizing transition
as the OL is weakened or the coefficient of nonlinearity is
decreased. The integrity of solitons is manifested by recovering
their original shape as the periodic potential or coefficient of
nonlinearity is restored to the initial value. This is in
agreement with the proposed physical mechanism associating the
delocalizing transition with quantum bound states in the effective
potential created by the soliton. As noted above, in 1D there are
always bound states in the confining potential, and that is why
the soliton doesn't disintegrate even when the depth of the
effective potential becomes infinitesimal.

\subsection{2D optical lattice}

We separately consider the delocalizing transition of bright solitons of
attractive and repulsive BECs in 2D OLs.
Starting point in thise cases, similarly to the previous situation, is
construction of a stationary solitonic state in a 2D OL.

\subsubsection{Attractive case}

\begin{figure}[htb]
\centerline{\includegraphics[width=8.0cm,height=6.0cm,clip]{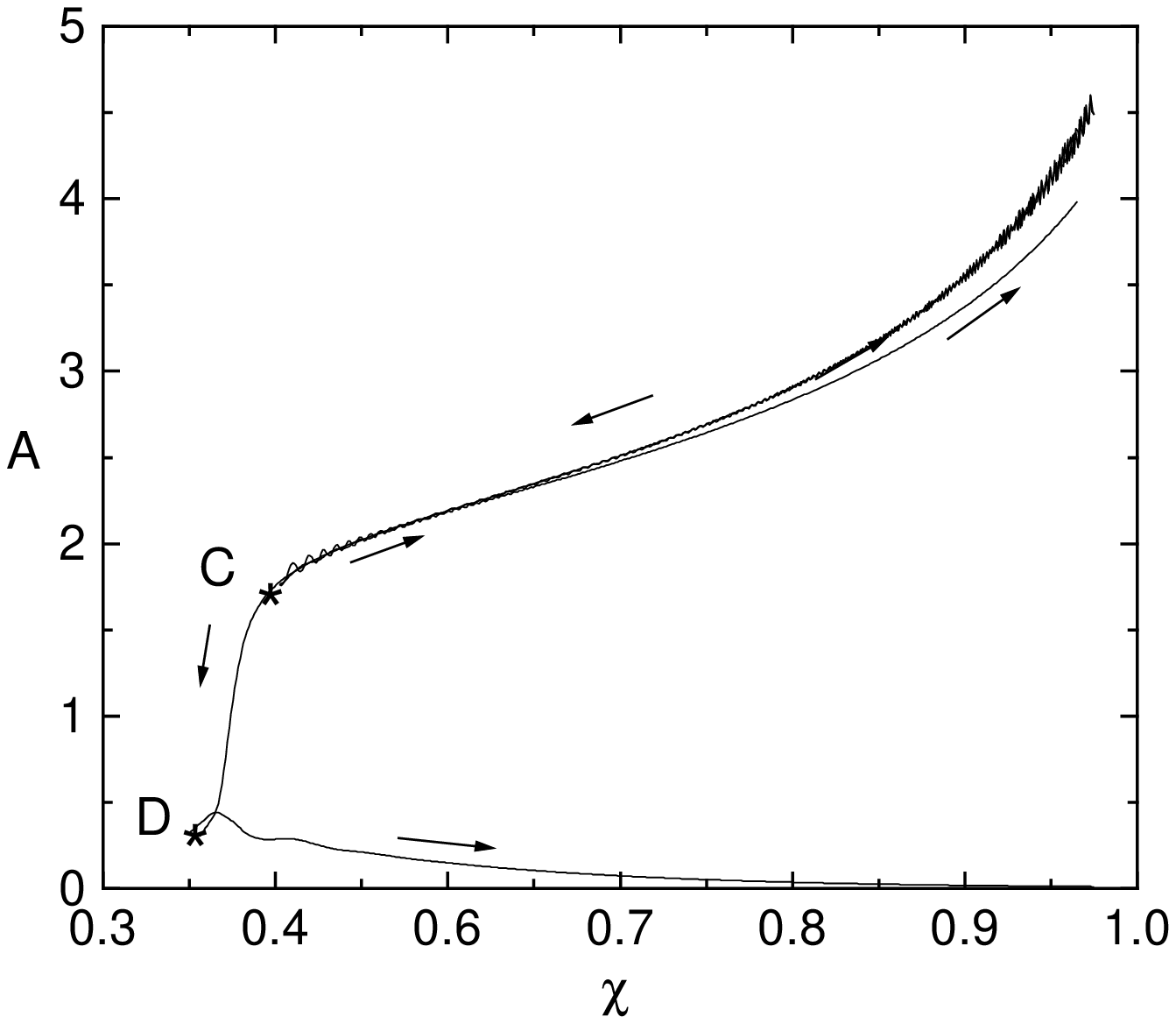}}
\caption{A 2D soliton of attractive BEC recovers its original amplitude
when the parameter $\chi(t)$ is decreased and then increased back without
crossing the critical value (near point C), and irreversibly disintegrates
when crossed (point D). A small deficit in the return value of the amplitude
is caused by the energy loss due to imperfect adiabaticity of the process.}
\label{f13}
\end{figure}

\begin{figure}[htb]
\centerline{\includegraphics[width=8.0cm,height=5.0cm,clip]{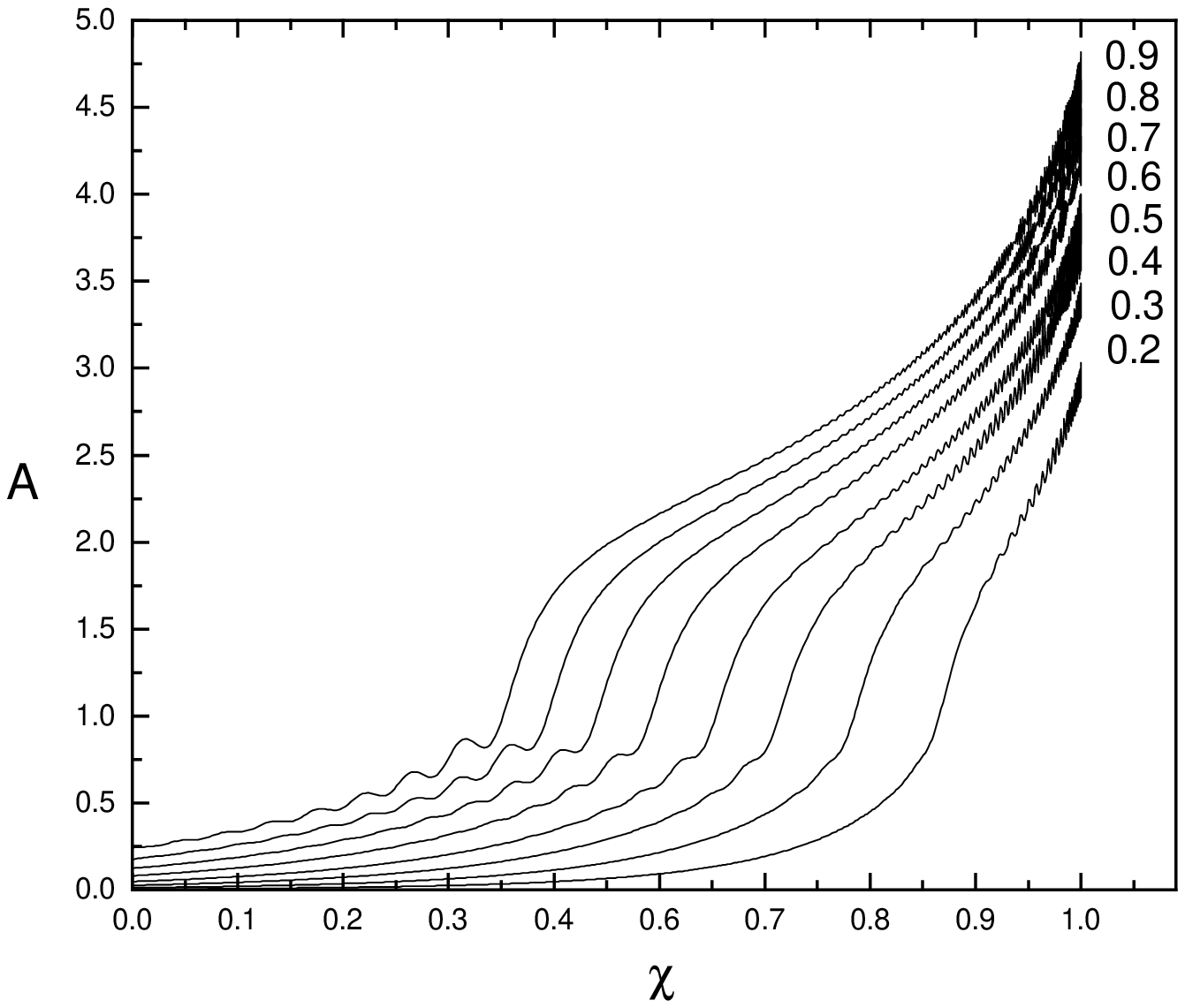}}
\vspace*{0.5cm}
\caption{Delocalizing transition of a 2D soliton in OLs of different
strength $\varepsilon$ (shown to the right of corresponding curves).}
\label{f14}
\end{figure}

\begin{figure}[htb]
\centerline{\includegraphics[width=8.0cm,height=6.0cm,clip]{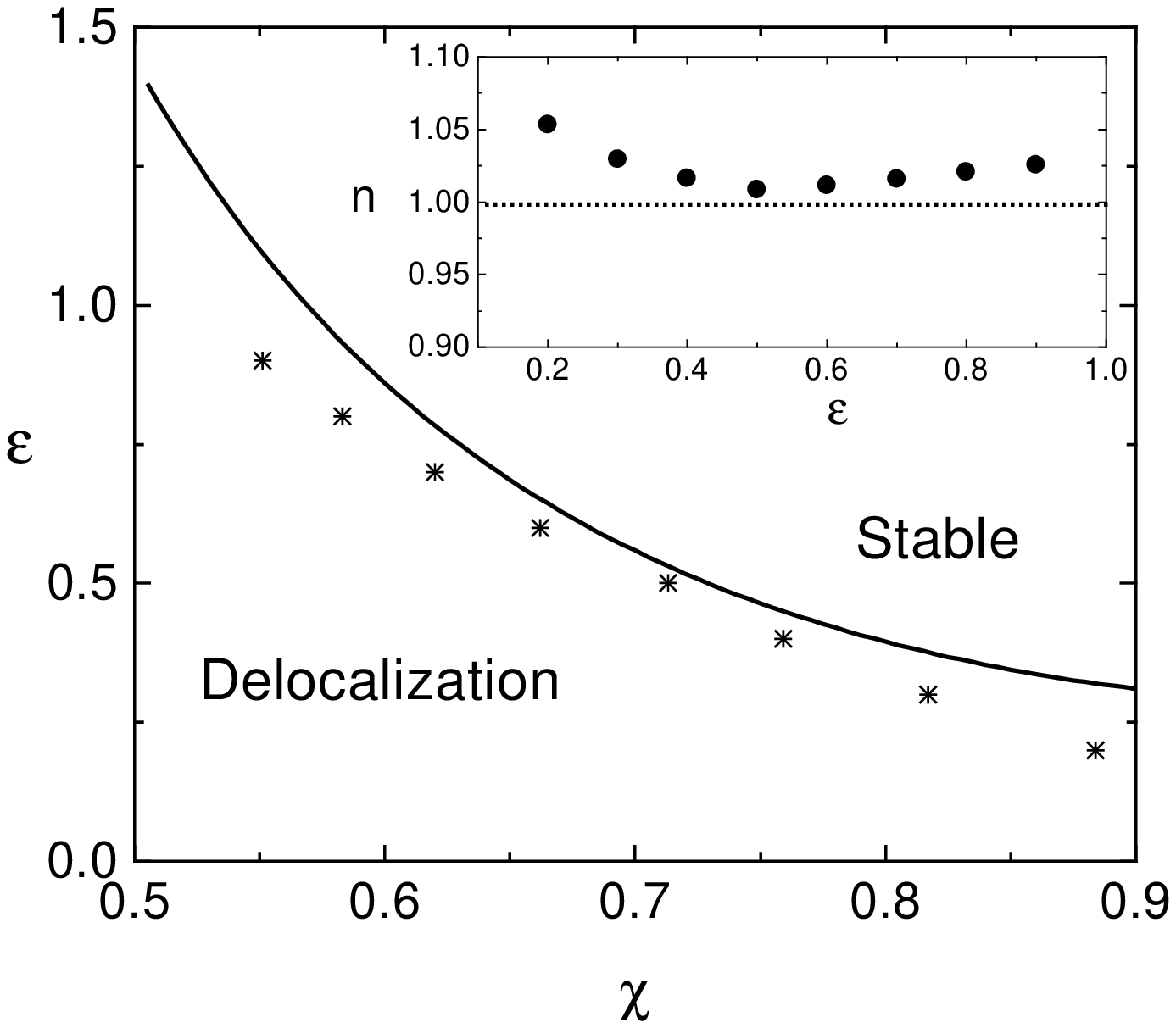}}
\vspace*{0.5cm}
\caption{The existence region of 2D solitons of attractive BEC.
Solid line - parametric solution of Eq.(\ref{nva}), also using the
conserved norm $N=\pi A_d^2/a_d$. Stars - delocalizing transition points
of Fig.\ref{f14}. The inset shows the number of quantum bound states
evaluated at the delocalizing transition points according to Eq.(\ref{n}).}
\label{f15}
\end{figure}

The stationary soliton of attractive BEC in 2D OL (Fig.\ref{f2}), generated
from the Gaussian waveform with parameters prescribed by the VA was
employed as initial condition in the GPE (\ref{gpe1}) with time-dependent
parameters $\varepsilon(t)$, or $\chi(t)$. By adiabatic variation of these
parameters one can bring the soliton close to the delocalizing transition
point, and then return back recovering the initial waveform, or induce
the delocalizing transition by slightly more decreasing the critical
parameter, as demonstrated in Fig.\ref{f13} by numerical solution of the
GPE (\ref{gpe1}).
Repeating similar delocalizing transition simulations one can establish
the existence region of 2D solitons in the parameter space
$\varepsilon$ vs. $\chi$. In Fig.\ref{f14} we present the result of
such numerical simulations for different strengths of the OL. Unlike the
Fig.\ref{f13}, the coefficient of nonlinearity $\chi(t)$ was linearly
decreased until zero, as we are interested in the critical values $\chi_d$
leading to delocalization of the soliton.
Fig.\ref{f15} represents the existence region for 2D solitons of attractive
BEC in the OL.
The number of quantum bound states in the effective potential
at the point of delocalizing transition, evaluated from
numerical simulations (Fig.\ref{f14}) and using the Eq.(\ref{n}),
appears to be very close to one (see the inset in Fig.\ref{f15})
for all values of $\varepsilon$, which supports the proposed physical model.

\subsubsection{Repulsive case}

\begin{figure}[htb]
\centerline{\includegraphics[width=8.0cm,height=4.0cm,clip]{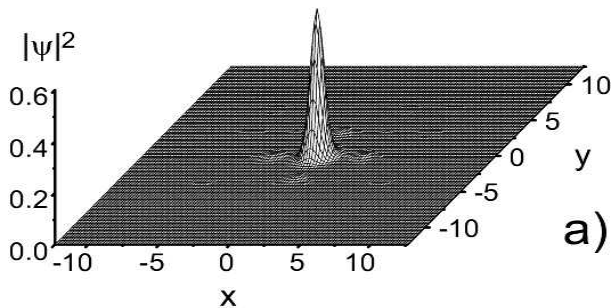}}
\centerline{\includegraphics[width=8.0cm,height=4.0cm,clip]{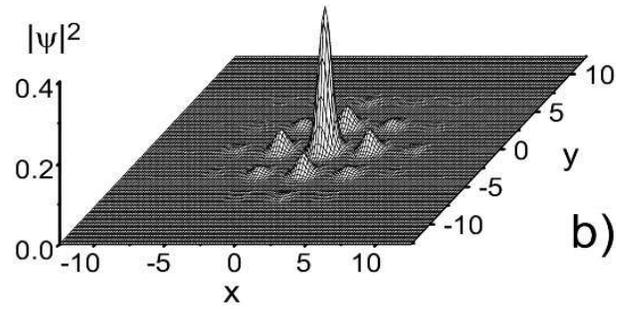}}
\vspace*{0.5cm}
\caption{a) A stable 2D soliton of repulsive BEC ($\chi = 1$)
in OL of strength $\varepsilon = 4.5$. b) The satellites
are more pronounced when the delocalizing transition point is approached by
reducing the coefficient of nonlinearity until $\chi = 0.4$.}
\label{f16}
\end{figure}

\begin{figure}[htb]
\centerline{\includegraphics[width=8.0cm,height=4.0cm,clip]{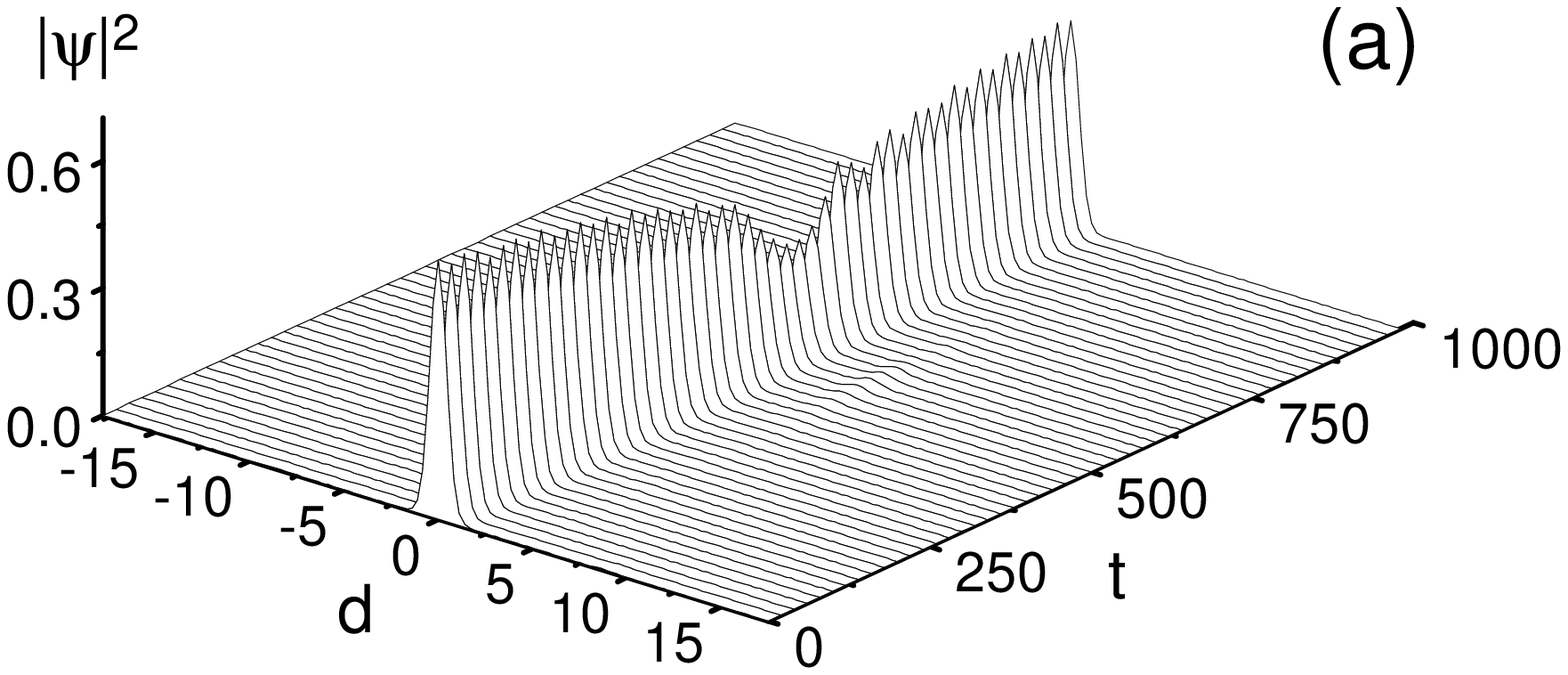}}
\centerline{\includegraphics[width=8.0cm,height=4.0cm,clip]{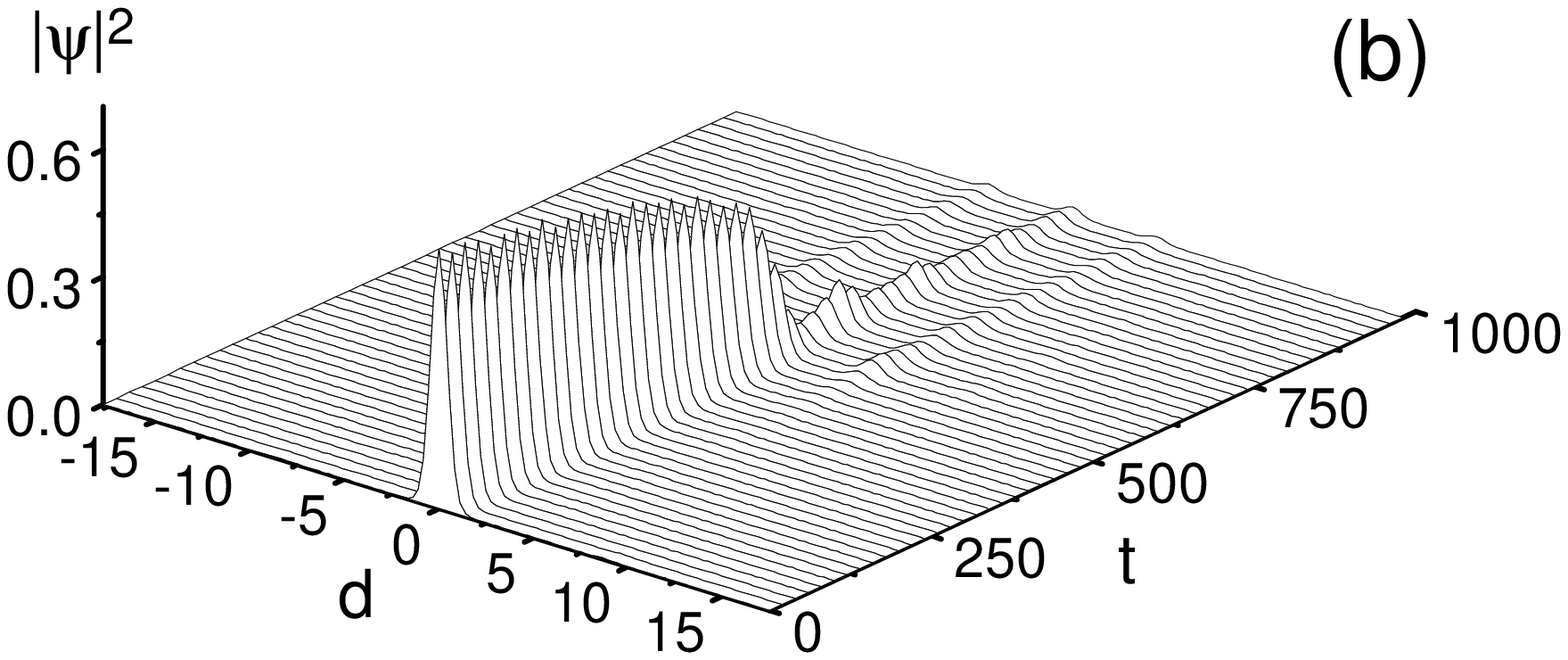}}
\vspace*{0.5cm}
\caption{Diagonal section profile for the time
evolution of a 2D soliton according to GPE (\ref{gpe1}). (a) The
waveform recovers its original shape when the magnitude of the
periodic potential is linearly decreased from $\varepsilon_0 = 4.5$
until $\varepsilon_{min} = 3.42$ at $t=500$, and then increased back
to $\varepsilon_0$ at $t=1000$. (b) Abrupt delocalizing transition occurs
as the strength of the OL is lowered below the critical value
$\varepsilon_c=  3.38$ at $t=500$. The Eq.(\ref{ft}) for variation of
$\varepsilon(t) = \varepsilon_0 \cdot f(t)$ is applied with $t_{end}=1000$,
and $\alpha=0.25$ for (a), $\alpha = 0.26$ for (b). }
\label{f17}
\end{figure}

As pointed out, matter-wave solitons of repulsive BEC in OLs
have a composite structure. The satellites are more pronounced when the
soliton is driven close to the delocalizing transition point in the
parametric space $\varepsilon$ vs. $\chi$, as illustrated in Fig.\ref{f16}.
Similarly to attractive case, the soliton can be brought close to the
delocalizing transition point by adiabatic change of parameters
$\varepsilon$ or $\chi$. Then, if the initial values of parameters
are restored without crossing the threshold, the soliton recovers its
original shape, otherwise it irreversibly disintegrates. Such a numerical
simulation with a soliton of repulsive BEC is presented in Fig.\ref{f17}.

\begin{figure}[htb]
\centerline{\includegraphics[width=8.0cm,height=4.0cm,clip]{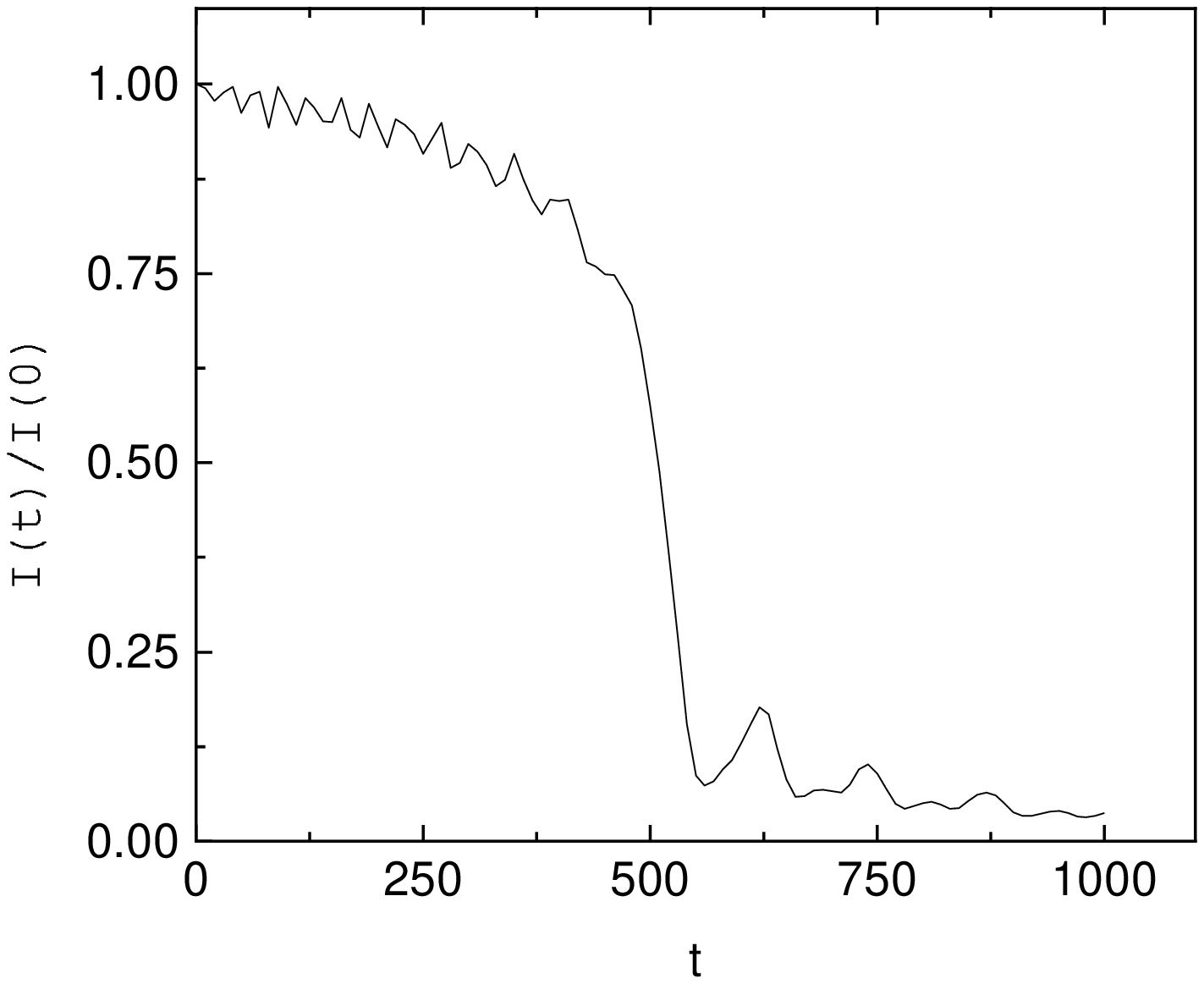}}
\vspace*{0.5cm}
\caption{Amount of BEC matter in the central peak of a 2D soliton
according to Eq.(\ref{int}), corresponding to Fig. \ref{f17}b. }
\label{f18}
\end{figure}

Of particular interest is the interplay between the central peak
and satellites of the lattice soliton in repulsive BEC. Some
information can be obtained by recording the amount of BEC matter $I(t)$
in the central peak (confined to a unit cell), while the parameters
$\varepsilon$ or $\chi$ adiabatically varied in time.
\begin{equation} \label{int}
  I(t)=\int_{-\pi/2}^{\pi/2} |\psi(x,y,t)|^2 dx dy.
\end{equation}
Time evolution of this quantity is presented in Fig.\ref{f18}

From this figure one can judge about the role of the central peak in the
integrity of the soliton. The delocalizing transition of a composite
soliton implies the rapid spreading of its central peak.

\subsection{3D optical lattice}

There is a growing interest in properties of BEC in 3D OLs
\cite{greiner2,greiner3}. The effect of dimensionality in
the process of delocalizing transition of matter-wave solitons is of
particular interest. Since we consider the phenomenon as rapid
spreading of the wave-packet through tunneling of BEC into
neighboring lattice sites, the main manifestation of the dimensionality
should be the increased sharpness of the transition (because of more
neighboring cells compared to 2D case).

\begin{figure}[htb]
\centerline{\includegraphics[width=8.0cm,height=4.0cm,clip]{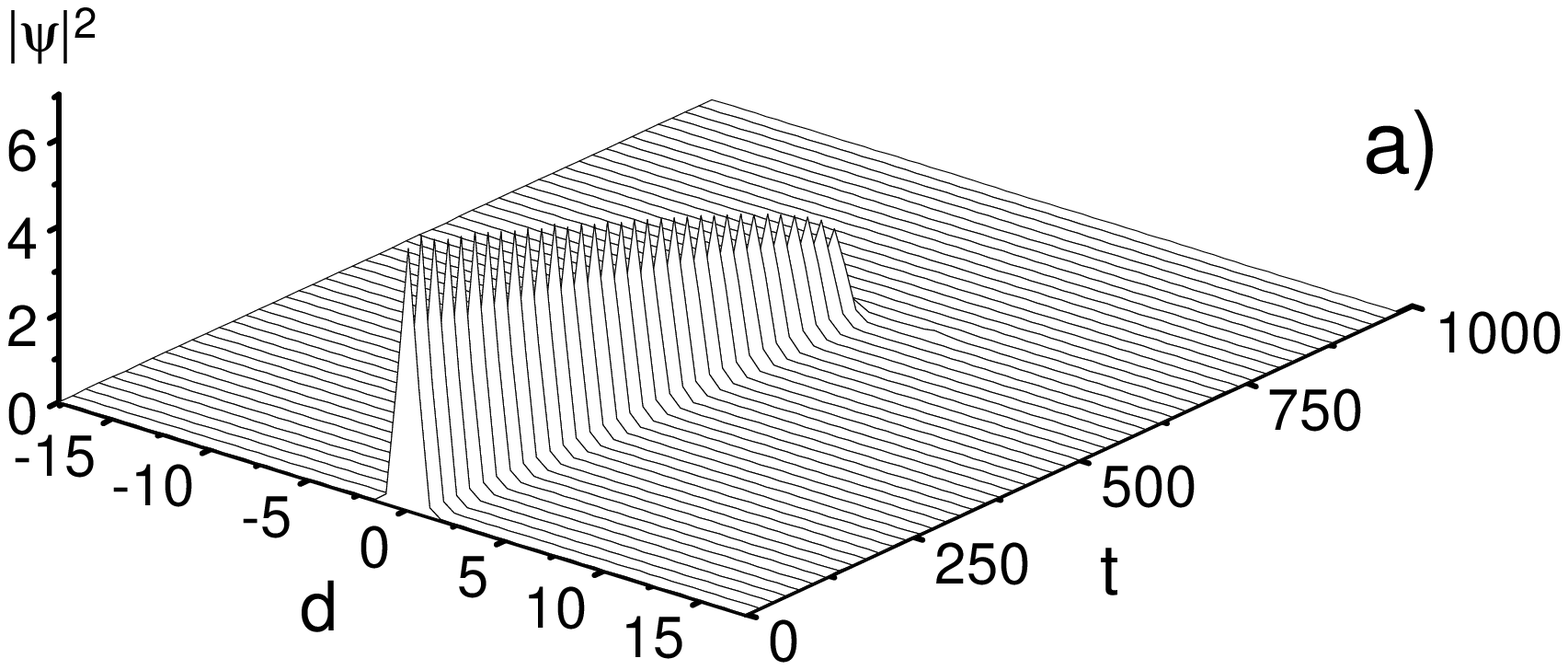}}
\vspace*{0.5cm}
\centerline{\includegraphics[width=8.0cm,height=4.0cm,clip]{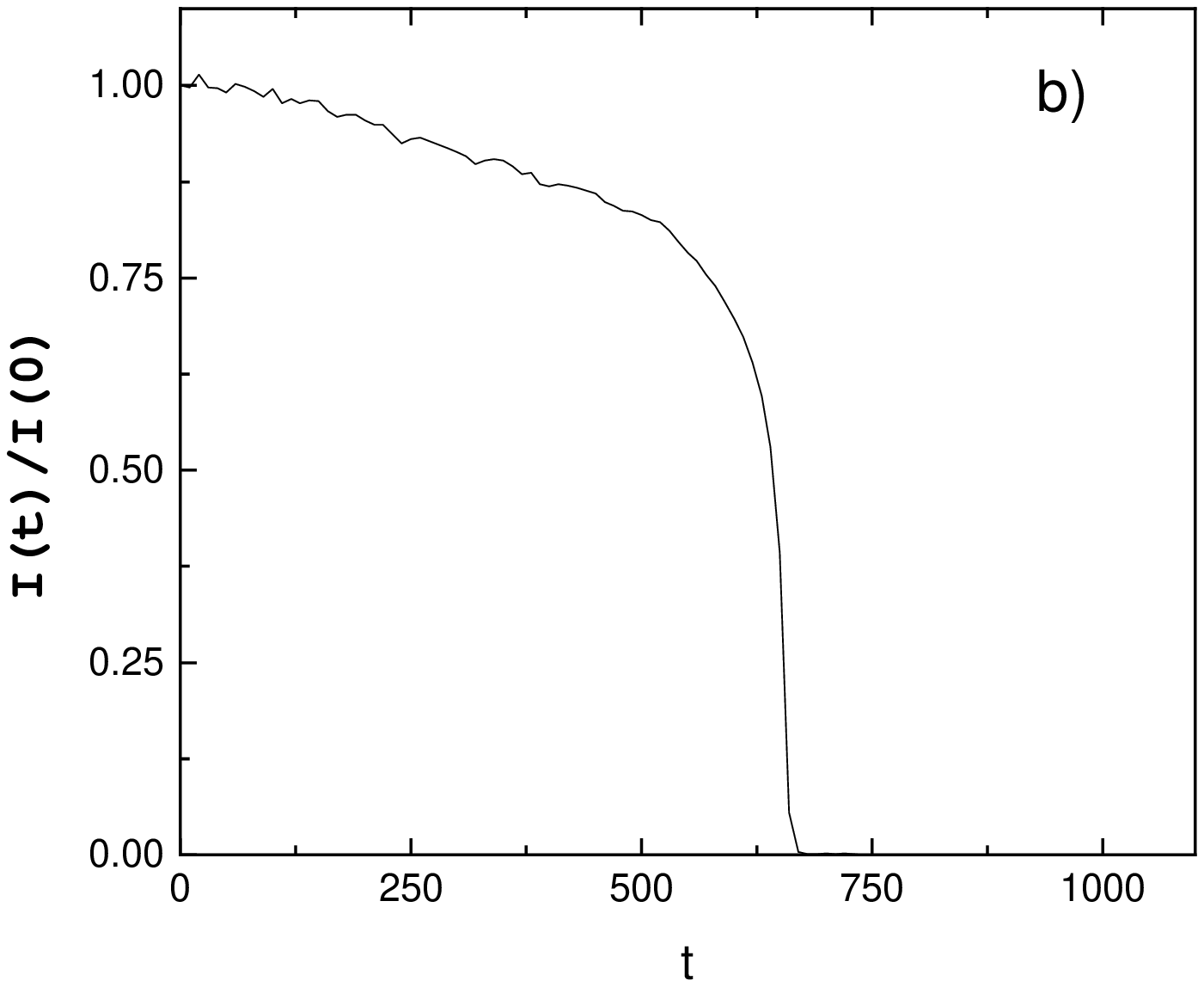}}
\vspace*{0.5cm}
\caption{Wave profile along the main diagonal of a cubic domain
$x,y,z \in [-4\pi, 4\pi]$ as obtained by numerical solution of the GPE
(\ref{gpe1}) with $\varepsilon(t) = \varepsilon_0 \cdot (1-t/t_{end})$ (a),
and corresponding integral Eq.(\ref{int}) over the central unit cell (b),
for a 3D OL with $\varepsilon_0 = 4.5, \ \chi = 1.0, \ t_{end} = 1000$.}
\label{f19}
\end{figure}

To verify the above conjecture, we performed numerical simulation of the
delocalizing transition of a soliton in 3D OL. The procedure for preparation
of a stable soliton and subsequent adiabatic variation of system parameters
are similar to 2D case. The result is presented in Fig.\ref{f19}.
As expected, the delocalizing transition in 3D appears to be more
sharp compared to 2D case.

Also note the absence of transient oscillations of the integral atomic
number after delocalizing transition in Fig.\ref{f19}b, as opposed to
2D case (Fig.\ref{f18}). This is another effect of higher dimensionality.

\section{Experimental feasibility}

Addressing the possibility of experimental observation of phenomena
considered in this paper, we note the rapid progress in manipulation
techniques for BEC in OLs \cite{greiner2,greiner3,denschlag}.
However, the matter-wave solitons in periodic potentials of OLs
have not been experimentally realized yet. Different approaches were
proposed, of which particular interest is the generation of gap solitons
in repulsive BEC employing the phenomenon of modulational
instability \cite{OL1D,bks2002}. Another possibility is the
independent successive formation of BEC in single sites of the OL as
proposed in Ref.\cite{burger2002}. The BEC initially occupying a single or few
lattice sites of the periodic potential quickly transforms into the solitonic
waveform (certainly, if the parameters are in the existence region),
as revealed from the numerical simulations.

After the matter-wave soliton is created, the delocalizing transition
can be induced by decreasing the strength of the OL through the
intensity of laser wave, or reducing the coefficient of nonlinearity
by changing the atomic scattering length via the Feshbach resonance.
The signature of the phenomenon is sudden disintegration of the
soliton at the critical point, therefore it will be accompanied by dropping
of the atomic density, which can be detected by imaging techniques. The
resolution of absorption-imaging systems currently in use
(which is $\sim 7 \, \mu$m \cite{burger2001}) is sufficient for controlling the
properties of BEC in the range of a unit cell of the OL.

Now let us estimate the dimensionless parameters in physical
units. In typical experiments to date the relevant parameters are
given by $n_0 = 10^{20}$ m$^{-3}$, $a_s =5.4$ nm, and $k_L =
2\pi/\lambda = 8.06 \cdot 10^6$ m$^{-1}$ for Rb \cite{burt}, and
$n_0 = 3 \cdot 10^{21}$ m$^{-3}$, $a_s =2.65$ nm, and $k_L = 1.07
\cdot 10^7$ m$^{-1}$ for Na \cite{inouye}. The strength of
nonlinear atomic interaction is given by $\chi = 8\pi n_0
a_s/k_L^2 = 0.21$ for Rb, and $\chi = 1.76$ for Na. Higher or
lower values of $\chi$ may be achieved by changing the density of
the condensate $n_0$, atomic scattering length $a_s$, or $k_L$.
All three parameters can be changed independently. The strengths
of OLs considered above are in the range of usual experimental
conditions \cite{morsch2003} $\varepsilon = 0 \div 20$ in units of
recoil energy $E_{rec} = \hbar^2 k_L^2/(2 m)$. Our numerical
simulation times $t \sim 1000$, well satisfying the adiabaticity
condition, correspond to $\sim 0.5$ s. The average lifetime of BEC
in optical trap, limited by three-body collisions and off-resonant
scattering of lattice photons is more than 3 s \cite{barrett}.
Therefore, observation of the delocalizing transition of solitons
is possible in the present experimental conditions.

\section{Conclusions}

We have studied the existence and delocalizing transition of
multidimensional solitons in periodic potentials by means of the
VA and direct numerical integration of the Gross-Pitaevskii equation.
VA provides the initial configuration for the soliton parameters which
can be used in PDE to generate a stable multidimensional soliton.

The most interesting property of multidimensional (2D and 3D) solitons
observed in this study is the delocalizing transition, which is manifested
as irreversible disintegration of the soliton at some critical strength of the
periodic potential, or coefficient of nonlinearity. Contrarily, 1D solitons
do not exhibit the delocalizing transition, retaining their integrity over
the whole range of parameter variations.

We proposed a physical mechanism for delocalizing transition of
solitons, according to which the existence of a soliton is
associated with the existence of quantum bound states in the effective
potential created by the soliton. At the point of delocalizing transition,
the effective potential has the critical strength to support only a single
bound state. As the strength becomes weaker than the critical value, the
effective potential cannot support a bound state, and as a consequence,
the soliton irreversibly disintegrates. Using the exactly solvable
P\"oschl-Teller potential as approximation for the effective potential
created by the soliton, we analytically determined the existence region
for 2D solitons. Numerical simulations of the Gross-Pitaevskii equation
have confirmed the validity of the proposed model.

Although we considered the problem with the emphasis on Bose-Einstein
condensates in OLs, multidimensional solitons and the phenomenon of
delocalizing transition can be observed in other relevant systems, e.g.
optically induced nonlinear waveguide arrays \cite{fleisher}.

\section*{Acknowledgements}
We aknowledge interesting discussions with  J. C. Eilbeck,
S. De Filippo, and B. A. Malomed. B. B. thanks the Physics
Department of the University of Salerno, Italy, for a two years
research grant during which this work was done. M. S. acknowledges
partial support from a MURST-PRIN-2000 Initiative, and from the
European grant LOCNET no. HPRN-CT-1999-00163.

\end{multicols}


\begin{references}
\bibitem[\dagger]{email} Permanent address: Physical-Technical Institute,
2-b, Mavlyanov str., 700084, Tashkent, Uzbekistan.

\bibitem{khaykovich}
L. Khaykovich {\it et al.,}
{\it Science} {\bf 296}, 1290, (2002);
K. E. Strecker {\it et al.,}
{\it Nature} {\bf 417}, 150, (2002).

\bibitem{burger}
S. Burger {\it et al.,}
Phys. Rev. Lett. {\bf 83}, 5198 (1999);
J. Denschlag {\it et al.,}
Science, {\bf 287}, 97 (2000).

\bibitem{carr}
L. D. Carr, Y. Castin, Phys. Rev. A {\bf 66}, 063602 (2002);
U. Al Khawaja {\it et al.,}
Phys. Rev. Lett. {\bf 89}, 200404 (2002);
V. Y. F. Leung {\it et al.,}
Phys. Rev. A {\bf 66}, 061602 (2002);
L. Salasnich, A. Parola, and L. Reatto, Phys. Rev. A {\bf 66}, 043603 (2002).

\bibitem{potting}
O. Zobay {\it et al.,} Phys. Rev. A {\bf 59}, 643 (1999);
S. P\"otting {\it et al.,}
J. Mod. Opt. {\bf 47}, 2653 (2000).

\bibitem{OL1D}
V. V. Konotop and M. Salerno, Phys. Rev. A {\bf 65}, 021602 (2002);
G. L. Alfimov, V. V. Konotop, and M. Salerno,
Europhys. Lett. {\bf 58}, 7 (2002);
K. M. Hilligs\oe, M. K. Oberthaler, and K. Marzlin,
Phys. Rev. A {\bf 66}, 063605 (2002);
G. L. Alfimov {\it et al.,} Phys. Rev. E {\bf 66}, 046608 (2002);
P. J. Y. Louis {\it et al.,}
Phys. Rev. A {\bf 67}, 013602 (2003).

\bibitem{bks2002}
B. B. Baizakov, V. V. Konotop, and M. Salerno,
J. Phys. B {\bf 35}, 5105 (2002).

\bibitem{ok2003}
E. A. Ostrovskaya and Y. S. Kivshar, e-print cond-mat/0303190.

\bibitem{bms2003}
B. B. Baizakov, B. A. Malomed, and M. Salerno,
Europhys. Lett. 2003 (submitted).

\bibitem{kalosakas}
G. Kalosakas, K. O. Rasmussen, and A. R. Bishop,
Phys. Rev. Lett. {\bf 89}, 030402 (2002).

\bibitem{jaksch}
D. Jaksch {\it et. al.,} Phys. Rev. Lett. {\bf 81}, 3108 (1998).

\bibitem{malomed99}
B. A. Malomed {\it et. al.,}
JOSA B {\bf 16}, 1197 (1999).

\bibitem{dalfovo}
F. Dalfovo, S. Giorgini, L. P. Pitaevskii, and S. Stringari,
Rev. Mod. Phys. {\bf 71}, 463 (1999).

\bibitem{anderson1983}
D. Anderson, Phys. Rev, {\bf A27}, 1393, 1983.

\bibitem{malomed2002}  B. A. Malomed, Progr. Optics {\bf 43}, 69 (2002).

\bibitem{abd2003}
F. Kh. Abdullaev, {\it et al.} Phys. Rev. A {\bf 67}, 013605 (2003).

\bibitem{berge}  L. Berg{\'{e}}, Phys. Rep. {\bf 303}, 260 (1998).

\bibitem{VK}
M. G. Vakhitov and A. A. Kolokolov, Radiophys. Quantum Electron.
{\bf 16}, 783 (1973).

\bibitem{numrecipes}  W. H. Press, S. A. Teukolsky, W. T. Vetterling, and B.
P. Flannery, {\em Numerical Recipes. The Art of Scientific Computing.}
(Cambridge University Press, 1996).

\bibitem{steel}
M. J. Steel and W. Zhang, e-print cond-mat/9810284;
H.Pu, {\it et al.}
Phys. Rev. A {\bf 67}, 043605 (2003).

\bibitem{poschl}
G. P\"oschl and E. Teller, Z. Phys. {\bf 83}, 1439 (1933).

\bibitem{flugge}
S. Fl\"ugge {\it Practical Quantum Mechanics}, Springer-Verlag,
1994.


\bibitem{greiner2}
M. Greiner {\it et al.}
Nature, {\bf 415}, 39 (2002).

\bibitem{greiner3}
M. Greiner {\it et al}
Nature, {\bf 419}, 51 (2002).

\bibitem{denschlag}
J. H. Denschlag {\it et al.}
J. Phys. B {\bf 35}, 3095, (2002).

\bibitem{burger2001}
S. Burger {\it et al.}
Phys. Rev. Lett. {\bf 86}, 4447 (2001).

\bibitem{burger2002}
S. Burger {\it et al.}
Europhys. Lett. {\bf 57}, 1 (2002).

\bibitem{burt}
E. A. Burt {\it et al.} Phys. Rev. Lett. {\bf 79}, 337 (1997).

\bibitem{inouye}
S. Inouye {\it et al.}
Nature, {\bf 392}, 151 (1998).


\bibitem{morsch2003}
O. Morsch {\it et al.}
Phys. Rev. A {\bf 67}, 031603 (2003).

\bibitem{barrett}
M. D. Barrett, J. A. Sauer, and M. S. Chapman,
Phys. Rev. Lett. {\bf 87}, 010404 (2001).

\bibitem{fleisher}
J. Fleisher {\it et al.}
Nature, {\bf 422}, 147 (2003);
J. Fleisher {\it et al.}
Phys. Rev. Lett. 90, 023902 (2003);


\end{references}
\end{document}